\documentclass[prd,nofootinbib,twocolumn,superscriptaddress,preprintnumbers,
balancelastpage,nofootinbib]{revtex4-1}

\usepackage{graphicx}  
\usepackage{dcolumn}   
\usepackage{bm}        
\usepackage{amssymb, amsmath}
\usepackage{slashed}   
\usepackage{color}

\usepackage[usenames,dvipsnames,svgnames,table]{xcolor}

\usepackage[colorlinks=true,linkcolor=Blue,citecolor=Blue,
urlcolor=Blue]{hyperref}

\usepackage[utf8]{inputenc}

\newcommand{\koto}{\textrm{KOTO}}
\newcommand{\na}{\textrm{NA62}}

\newcommand{\vtx}{\textrm{vtx}}
\newcommand{\ses}{\textrm{SES}}

\newcommand{\MeV}{\textrm{MeV}}
\newcommand{\GeV}{\textrm{GeV}}

\newcommand{\cL}{\mathcal{L}}

\newcommand{\BR}{\mathcal{B}}

\begin{document}
\widetext

\title{Imprint of a new light particle at KOTO?}

\author{Yi Liao}
\email{liaoy@nankai.edu.cn}
\affiliation{School of Physics, Nankai University, Tianjin 300071, China}
\affiliation{Center for High Energy Physics, Peking University, Beijing 100871, China}

\author{Hao-Lin Wang}
\email{whaolin@mail.nankai.edu.cn}
\affiliation{School of Physics, Nankai University, Tianjin 300071, China}

\author{Chang-Yuan Yao}
\email{yaocy@nankai.edu.cn}
\affiliation{School of Physics, Nankai University, Tianjin 300071, China}

\author{Jian Zhang}
\email{zhangjianphy@nankai.edu.cn}
\affiliation{School of Physics, Nankai University, Tianjin 300071, China}

\begin{abstract}
\noindent
Recently, the KOTO experiment reported their new preliminary result of searching for the decay $K_L\to\pi^0\nu\bar{\nu}$. Three candidate events were observed in the signal region, which exceed significantly the expectation based on the standard model. On the other hand, the new NA62 and previous BNL-E949 experiments yielded a consistent result and confirmed the standard model prediction in the charged meson decay $K^+\to\pi^+\nu\bar{\nu}$. Furthermore, the two decays are bound by a well-motivated relation from an analysis of isospin symmetry that is hard to break by the new physics of heavy particles. In this work, we study the issue by a systematic effective field theory approach with three of the simplest scenarios, in which the $K_L$ may decay into a new light neutral particle $X$, i.e., $K_L\to\pi^0X$, $K_L\to \gamma\gamma X$, or $K_L\to\pi^0XX$. We assess the feasibility of the scenarios by simulations and by incorporating constraints coming from NA62 and other relevant experiments. Our main conclusion is that the scenario $K\to\pi XX$ for a long lived scalar $X$ seems more credible than the other two when combining distributions and other experimental constraints while the region below the KOTO's blind box provides a good detection environment to search for all three scenarios for a relatively heavy $X$.
\end{abstract}


\maketitle

\section{Introduction}
\label{sec:intro}

The flavor changing neutral current decays of the neutral and charged kaons $K \to \pi \nu \bar{\nu}$ provide a clean venue to examine precisely the standard model (SM) and to search for new physics beyond it. Recently, the KOTO experiment reported their preliminary result for the decay $K_L\to\pi^0\nu\bar{\nu}$ using data collected during years $2016-2018$~\cite{kototalks}. With the KOTO single event singularity (SES), $\ses_{\pi^0 \nu \bar{\nu}}= 6.9 \times 10^{-10}$, which is larger than the SM prediction $\BR\left( K_L \to \pi^0 \nu \bar{\nu} \right)_{\rm SM \ \, } = \left( 0.34 \pm 0.06 \right) \times 10^{-10}$~\cite{Buras:2015qea}, a total of $0.10\pm 0.02$ events would be expected. But surprisingly, three candidate events apparently fell into the signal region. As the experimental analysis is still in progress, any theoretical attempt to interpreting the potential discrepancy has to be taken with great caution. On the other hand, if the three candidate events were confirmed in the future, it would imply a decay branching ratio~\cite{Kitahara:2019lws}:
\begin{eqnarray}
\BR\left( K_L \to \pi^0 \nu \bar{\nu} \right)_\koto
= 2.1^{+2.0 (+4.1)}_{-1.1 (-1.7)} \times 10^{-9},
\end{eqnarray}
at 68\% (95\%) C.L., and would pose a strong challenge to the SM.

In the charged sector, the NA62 experiment recently reported their result~\cite{NA62talks}:
\begin{eqnarray}
\BR\left( K^+ \to \pi^+ \nu \bar{\nu} \right)_\na
= \left( 0.47^{+0.72}_{-0.47} \right) \times 10^{-10},
\end{eqnarray}
which is consistent with a previous measurement by the BNL-E949 Collaboration, $\BR\left(K^+\to\pi^+\nu\bar{\nu}\right)_\textrm{E949}= \left(1.73^{+1.15}_{-1.05}\right)\times 10^{-10}$~\cite{Artamonov:2008qb}. Combining the two one obtains
\begin{eqnarray}
\BR\left( K^+ \to \pi^+ \nu \bar{\nu} \right)_{\rm Exp} &=& \left( 0.86^{+0.62}_{-0.58} \right) \times 10^{-10},
\label{eq:expk+}
\end{eqnarray}
or an upper bound at 95\% C.L.,
\begin{eqnarray}
\BR\left( K^+ \to \pi^+ \nu \bar{\nu} \right)_{\rm Exp}
&\leq& 2.05 \times 10^{-10}.
\label{eq:expk+95}
\end{eqnarray}
This confirms the SM prediction $\BR\left( K^+ \to \pi^+ \nu \bar{\nu} \right)_{\rm SM \;} = \left( 0.84 \pm 0.10 \right) \times 10^{-10}$~\cite{Buras:2015qea}.

This potential tension between the neutral and charged sectors is even exacerbated due to a theoretical relation between the two. Based on the well-motivated assumption that the decays are dominated by the  interactions with isospin change $\Delta I=1/2$, they are related by the so-called Grossman-Nir (GN) bound~\cite{Grossman:1997sk},
\begin{eqnarray}
  \BR(K_L\to \pi^0 \nu \bar{\nu}) \lesssim 4.3 \BR(K^+\to\pi^+  \nu \bar{\nu}).
  \label{eq:GNb}
\end{eqnarray}
Together with the bound in the charged sector of equation~\eqref{eq:expk+95}, this implies $\BR\left( K_L \to \pi^0 \nu \bar{\nu} \right) < 8.8 \times 10^{-10}$ at $95\%$ C.L., which would lead to at most 1.3 instead of 3 candidate events at KOTO. This bound is hard to break as it is immune to low energy effects at leading order of new heavy particles, and may only be violated by $\Delta I=3/2$ interactions of higher dimensional operators between quarks and neutrinos; for recent discussions and attempts, see, for instance, Refs.~\cite{Li:2019fhz,He:2020jzn,He:2020jly,Ziegler:2020ize}.

The challenge here is to accommodate the KOTO result while respecting the measurements at NA62 and others. In this work we propose to relax the tension by assuming a light neutral particle that can appear in the final states of kaon decays. We will consider three scenarios, one of which has been suggested in the literature, and will assess by simulations whether they are practically feasible. Since $\pi^0$ decays into a pair of photons and the neutrino pair appears as missing energy, there are three simplest scenarios that could mimic the SM decay searched for at KOTO, namely, $K_L\to\pi^0X$, $K_L\to\gamma\gamma X$, and $K_L\to\pi^0XX$, in which $X$ appears as missing energy.

We will first examine in Sec.~\ref{sec.2} the most popular scenario $K \to\pi X$~\cite{Kitahara:2019lws,Fuyuto:2014cya, Fuyuto:2015gmk, Egana-Ugrinovic:2019wzj,Dev:2019hho,Jho:2020jsa,Liu:2020qgx,Cline:2020mdt}. We will assess whether it is feasible by simulating its event distribution and comparing it with that measured at KOTO. We will also incorporate constraints coming from other measurements on $K_L$ and $K^+$ decays. In Sec.~\ref{sec.3} we consider the loop-induced process of $K_L\to\gamma\gamma X$ which could disguise itself as a signal event for the decay searched for at KOTO but has no counterpart at NA62. If the new particle couples only in pairs to quarks for one reason or another, it will also appear in pairs in the kaon decays. We will therefore make an analysis on the decays $K\to\pi XX$ in Sec.~\ref{sec.4}. As we are conscious of the preliminary status of the KOTO result, we will show what constraint could be set on the $K_L$ decay branching ratio if any of the three candidate events were confirmed or completely falsified in the future. In the course of our analysis we will make some suggestions that may deserve further study in the KOTO and NA62 experiments. We summarize our main results in the last section~\ref{sec.5}. Our analysis in the following sections is based on the effective field theory approach. The effective interactions at leading order and the amplitudes and branching ratios for various decays are detailed in Appendixes~\ref{EFT}, \ref{Detailedcalc}, and \ref{Relevant results}, and the simulation setup and validation is discussed in Appendixes~\ref{Simulation}. We mention for completeness that if there are new light particles other than the $X$ particle that mediate the decays considered here, they could break the GN bound in a different manner or change the decay kinematical distributions to avoid the GN bound. We refer the interested reader to Refs.~\cite{Ziegler:2020ize,Fabbrichesi:2019bmo} on this approach.

\section{$K_L\to\pi^0X$}
\label{sec.2}

\begin{figure*}[!t]
  \hspace{-0.5cm}

\includegraphics[width=0.42\textwidth]{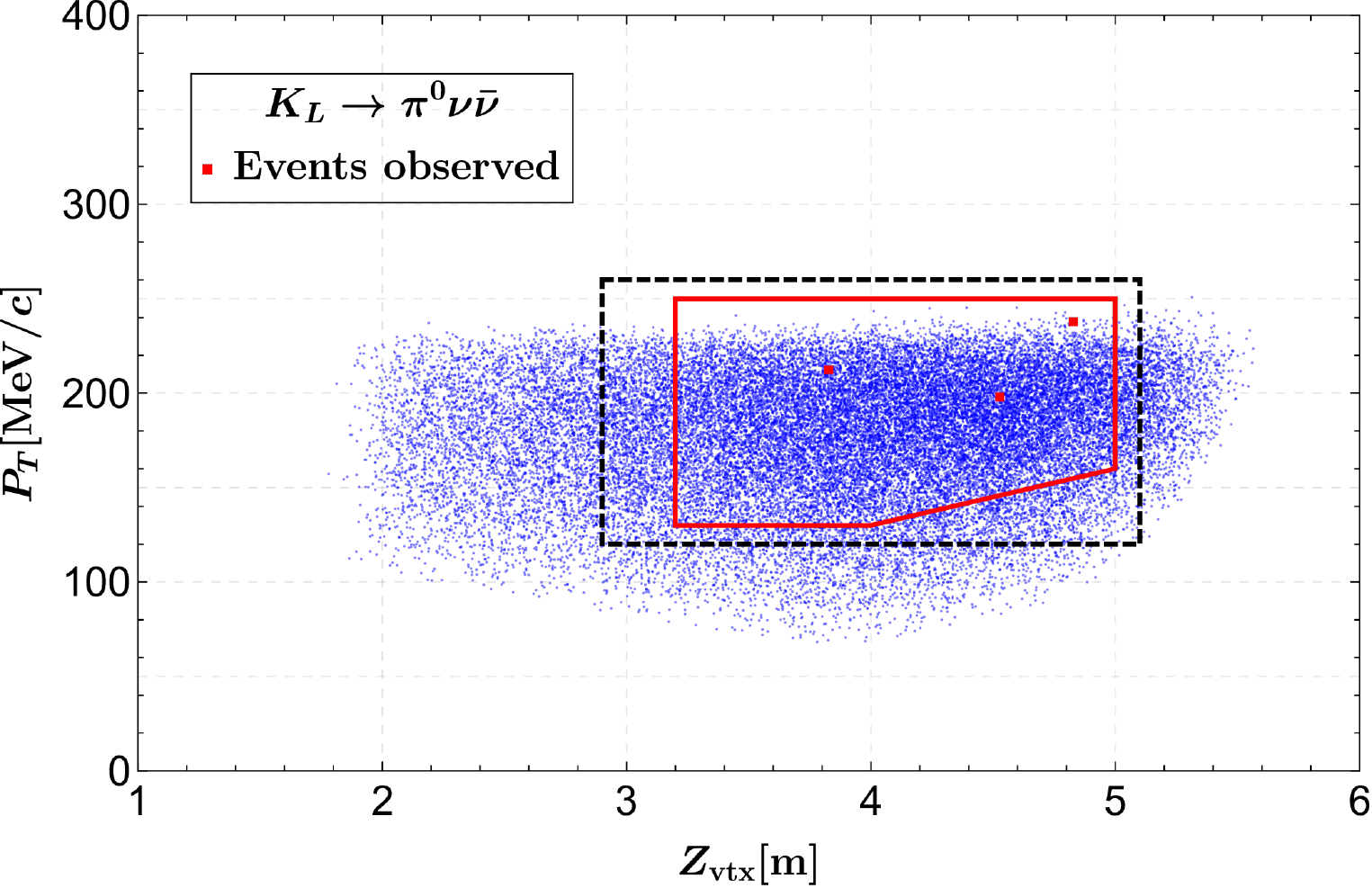}
\includegraphics[width=0.42\textwidth]{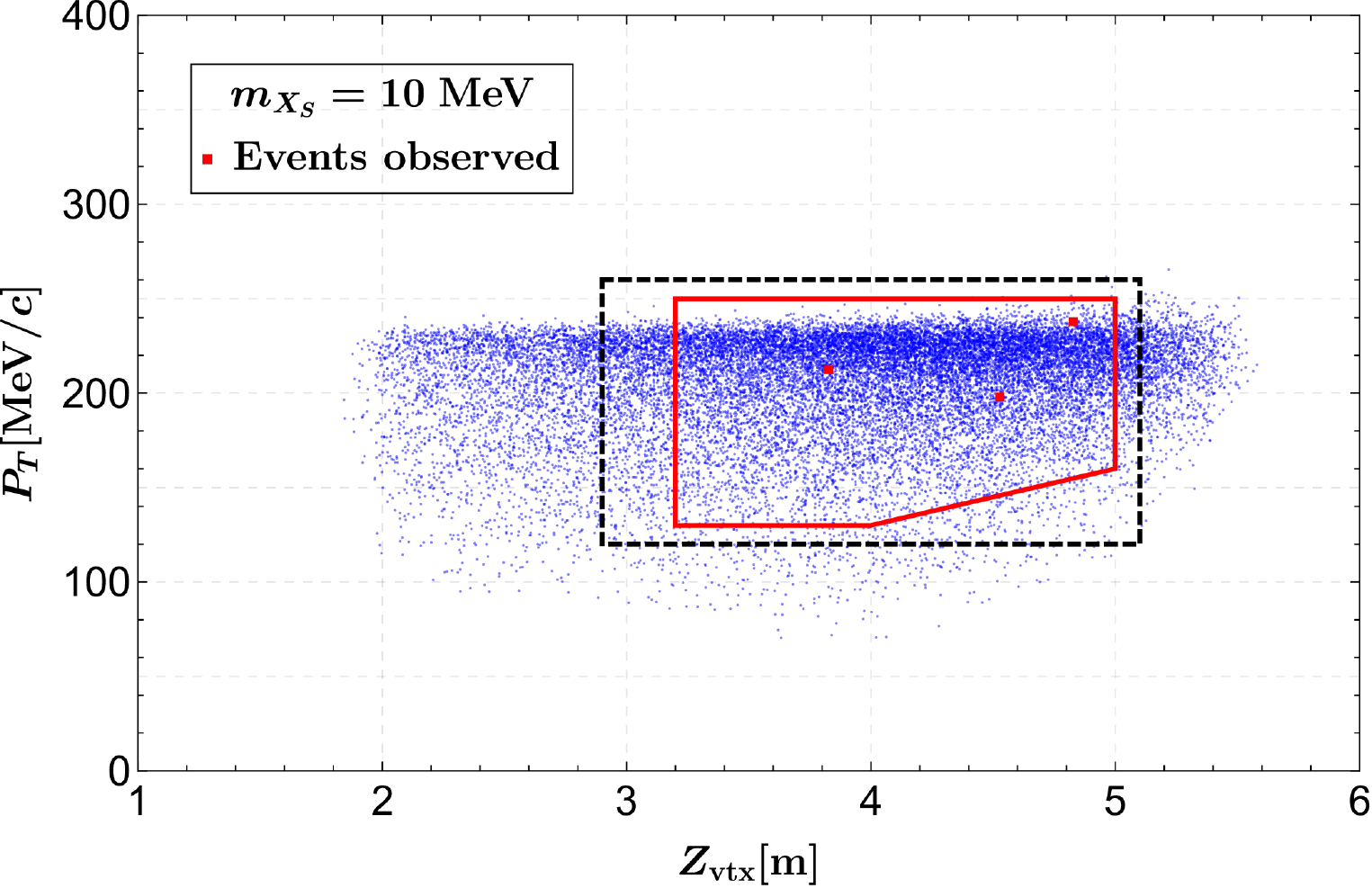} 
\includegraphics[width=0.42\textwidth]{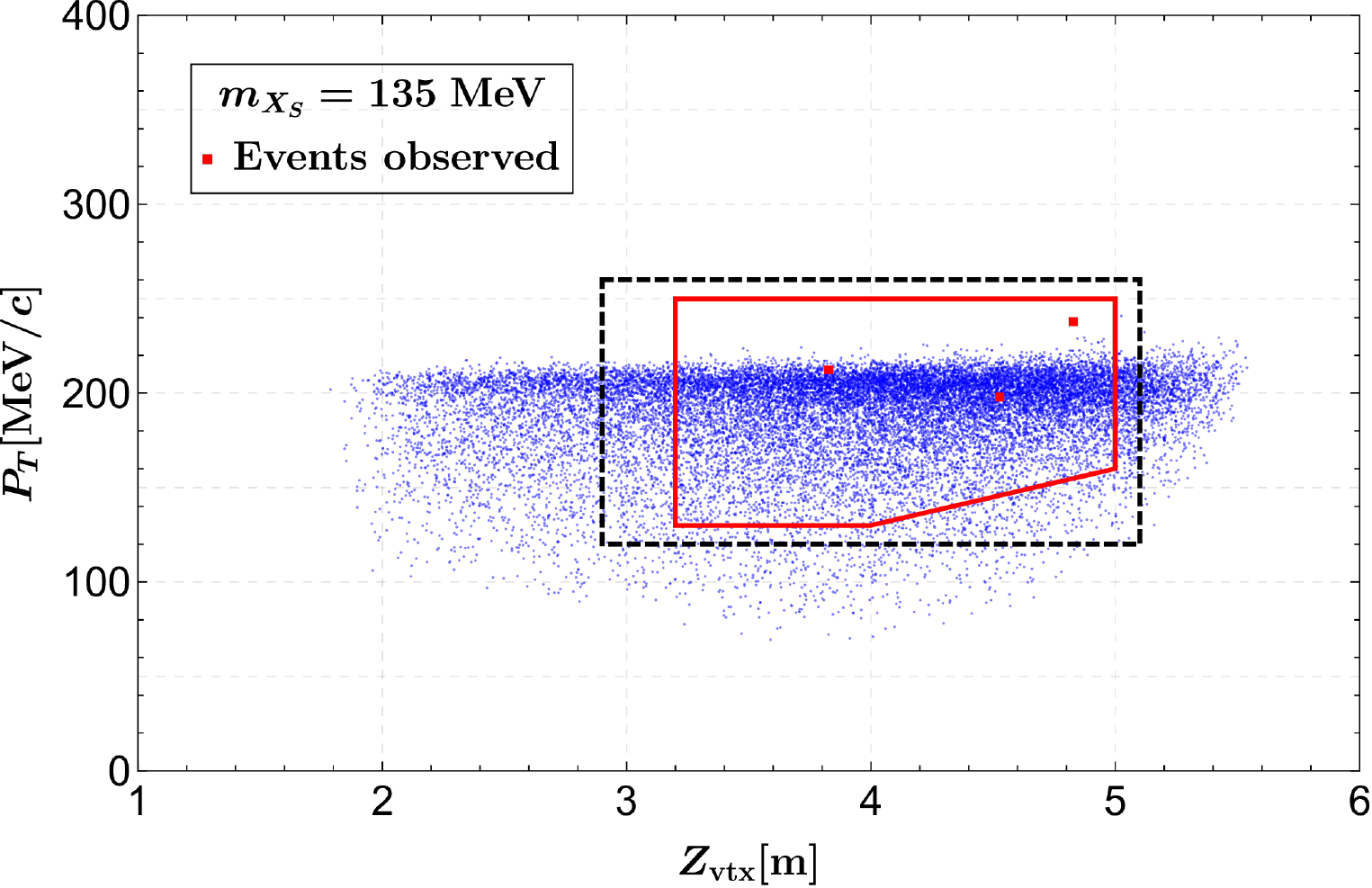} 
\includegraphics[width=0.42\textwidth]{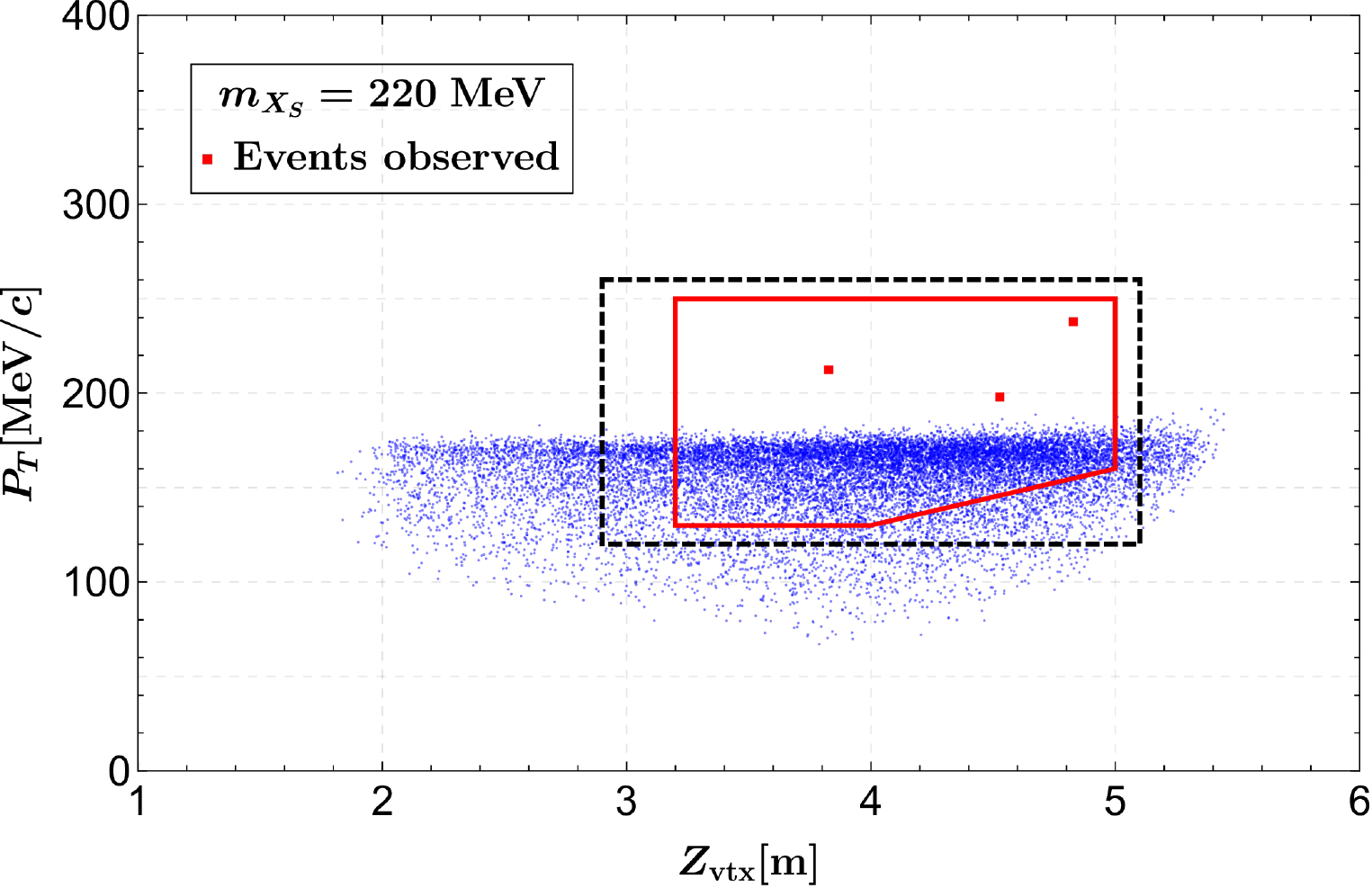} 

\caption{Reconstructed events (blue scatters) in the $Z_\vtx-P_T$ plane after all the cuts for $K_L\to\pi^0\nu\bar{\nu}$ (top left) and for $K_L\to\pi^0X_S$ with $m_{X_S} = 10~\MeV$ (top right), $135~\MeV$ (bottom left), and $220~\MeV$ (bottom right). The results for a vector $X_V$ are the same, as shown in Eqs.~\eqref{eq:distpixs} and \eqref{eq:distpixv}. The region surrounded by the red solid (black dashed) lines is the signal region (blind box) implemented in the KOTO analysis~\cite{kototalks} which yielded the three candidate events in red squares.}
\label{fig:eventdistr}
\end{figure*}

The signal region for the $K_L\to\pi^0\nu\bar{\nu}$ search at KOTO is also suitable for the two-body decay $K_L\to\pi^0X$, where $X$ acts as missing energy. Assuming parity and working to leading order in low energy expansion, $X$ can be a scalar or vector particle. However, as shown in Fig.~\ref{fig:eventdistr}, the signal distributions in the $Z_\vtx-P_T$ plane of the two decays are different. Here $Z_\vtx$ is the $\pi^0$ decay vertex position projected onto the $K_L$ beam direction, and $P_T$ is its transverse momentum with respect to the beam. We have performed the simulations by applying the kinematical cuts proposed in KOTO's 2015 data analysis~\cite{Ahn:2018mvc} and assuming the signal region chosen from its $2016-2018$ data analysis~\cite{kototalks}. While the decay $K_L\to\pi^0\nu\bar{\nu}$ has a relatively uniform distribution in the signal region, $K_L\to\pi^0X$ concentrates in a narrow interval of $P_T$. Furthermore, the heavier the $X$ is, the lower the maximal $P_T$ is. For a heavy $X$, it is hard to accommodate the candidate events with a high $P_T$ at KOTO.

\begin{figure}[t!]
\hspace{-0.5cm}
\includegraphics[width=0.42\textwidth]{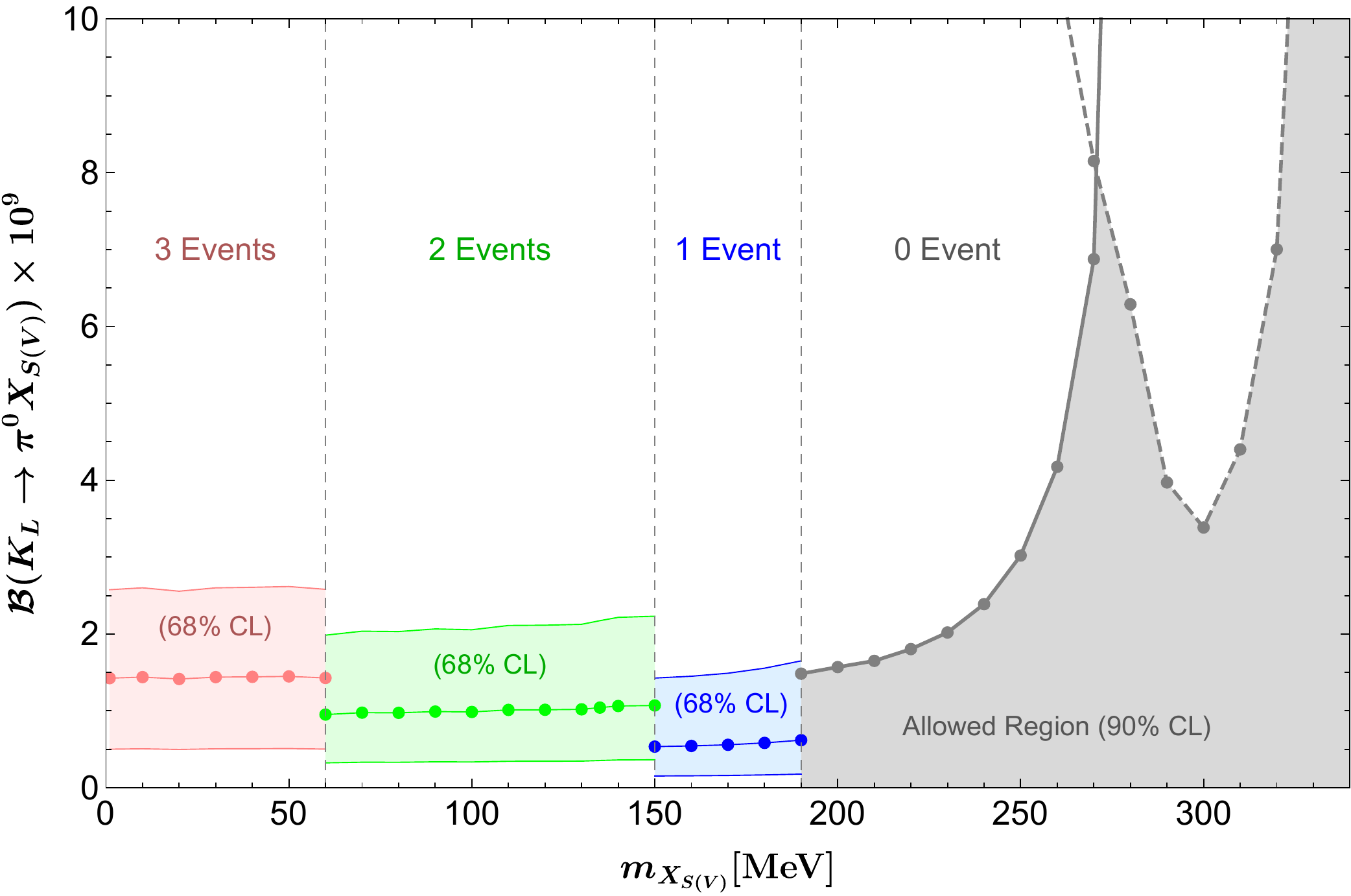}
\caption{Branching ratio for $K_L\to\pi^0X_{S(V)}$ with 68\% C.L. limits in various mass intervals that can accommodate a specific number of candidate events, and its 90\% C.L. upper limit as a function of mass in the KOTO signal region (solid line) and in the region below the blind box (dashed line). The dots correspond to the points $m_{X_{S(V)}}=1,~10,~20,\dots,~130,~135,~140,\dots,~290,~300~\MeV$.}
\label{fig:brdistriXS}
\end{figure}

The branching ratio for $K_L\to\pi^0X$ corresponding to a specific number of candidate events, $N_{\rm signal}$, that can be accommodated is estimated by the relation
\begin{eqnarray}
&&\BR\left( K_L \to \pi^0 X \right)_{\rm KOTO}
\nonumber
\\
&=& N_{\rm signal} \cdot {\rm SES}_{\pi^0X}
\label{BRKOTO}
\\
&=& N_{\rm signal} \cdot {\rm SES}_{\pi^0\nu \bar{\nu}} \cdot \frac{\epsilon_{\pi^0\nu \bar{\nu}}}{\epsilon_{\pi^0X}(m_{X})}.
\end{eqnarray}
Here $\epsilon$ is the detection efficiency in the signal region at KOTO~\cite{kototalks} upon imposing various kinematical cuts~\cite{Ahn:2018mvc}. We have estimated the ratio of the two efficiencies by simulations and then used the above quoted $\ses_{\pi^0 \nu \bar{\nu}}$ (with an assumed relative error of $10\%$) to obtain $\ses_{\pi^0X}$. The total background is $0.10\pm 0.02$ in the signal region (about one half from the decay $K_L \to \pi^0 \nu \bar{\nu}$ and another half from its SM background) and $0.08 \pm 0.06$ in the region below the blind box~\cite{kototalks}. The branching ratio for $K_L\to\pi^0X$ with two-sided 68\% C.L. limits is shown in Fig.~\ref{fig:brdistriXS} as a function of $m_X$ in various mass intervals where a specific number $N_{\rm signal}$ of candidate events are accommodated. Note that the choice of the interval delimiters is not sharp but only a rough estimate based on simulations. When $X$ is so heavy (roughly $m_X>190~\MeV$) that none of the candidate events can be accommodated or even all of its signals drop below the blind box ($m_X>270~\MeV$), we then use the above background estimates to set a 90\% C.L. upper bound on the decay branching ratio.

Now we examine whether the decay $K_L \to \pi^0 X$ offers a feasible interpretation to the candidate events observed in the KOTO signal region by a comprehensive analysis of the decay distribution and the limits set by other experiments.

We consider first the $X$ mass intervals $m_X\in(100,165)\cup(260,354)~\MeV$. These intervals were not taken into account in the NA62 experiment since its sensitivity to the decay $K^+\to\pi^+X$ is considerably degraded by the large backgrounds $K^+\to\pi^+\pi^0(\gamma)$ and $K^+\to\pi^+\pi^0\pi^0,~\pi^+\pi^+\pi^-$, respectively. Thus the restrictive GN bound is practically not in action. Nevertheless, the second interval is obviously not supported by the KOTO signals as seen in Fig.~\ref{fig:eventdistr} and the first one cannot provide a perfect solution either. For $X$ in the first interval, i.e., $m_X \sim m_{\pi^0}$, the decay $K_L\to\pi^0X$ can only accommodate two candidate events but not the one with a high $P_T \sim 238~\MeV$. If $X$ is stable and invisible and if we leave aside the last event, we may obtain the branching ratio with two-sided limits at $68\%$ ($95\%$) C.L. at, e.g., $m_X=135~\MeV$,
\begin{eqnarray}
\BR\left( K_L \to \pi^0 X \right)_{\rm KOTO} = 1.12^{+1.21(+2.70)}_{-0.74(-0.97)} \times 10^{-9}.
\label{eq:X-KOTO}
\end{eqnarray}
It might be appropriate to recall that a stable and invisible $X$ has been studied as a light dark matter particle that appears singly or in a pair in rare kaon to pion decays; see, for instance, Refs.~\cite{Pospelov:2017pbt,Pospelov:2007mp,Bezrukov:2009yw}. But such a stable particle would hardly relax the tension between the NA62 and preliminary KOTO results due to the GN bound. Our strategy is to utilize their different experimental settings so that $X$ escapes the KOTO detector but decays in the NA62 detector without breaking the GN bound. The only potential loophole may appear in the region around $m_X\sim m_\pi$ which has been cut in the NA62 experiment. But as analyzed above, even this fortuitous situation does not work out perfectly.

For $X$ of the other mass, it must be so long lived to be invisible at KOTO while short lived to decay into SM particles to be vetoed at NA62, thus avoiding the constraint from the GN bound. In order to incorporate all three candidate events, $X$ must be light enough, $m_X \lesssim 60~\MeV$, but still it is difficult to offer a feasible solution. First of all, not all observed events are gracefully in the main distribution region as can be seen in Fig.~\ref{fig:eventdistr}. More importantly, as we will detail below, this light mass region is tightly constrained by other experiments.

For $m_X\lesssim 60~\MeV$, the signals of $K^+\to\pi^+X$ at NA62 would be predominantly located in its signal region 1 for the $K^+\to\pi^+\nu\bar\nu$ search~\cite{NA62talks} with missing mass squared $m_{\rm miss}^2 \in (0, 0.01)~\GeV^2$, where no events were observed. Since we did not simulate the decays $K^+\to\pi^+ X,~\pi^+\nu\bar\nu$ at NA62, we attempted an estimate on some simplifying assumptions and by employing its combined 2016 and 2017 data. We assumed an equal $\ses=(0.346 \pm 0.017)\times 10^{-10}$~\cite{NA62talks} for both decays and an $m_X$-independent efficiency. Utilizing the expected signals ($0.81\pm 0.10$) and background ($0.60\pm 0.06$) for $K^+\to\pi^+\nu\bar{\nu}$ corresponding to their total expectations in signal regions 1 and 2 at NA62~\cite{NA62talks} and their distributions~\cite{Marchevski:2019ioc}, we obtained $\BR\left( K^+ \to \pi^+ X \right)_{\rm NA62}< 0.4 \times 10^{-10}$ at $90\%$ C.L.. Very recently, the NA62 group presented their preliminary result through systematic simulations based on the 2017 data, and set a limit $\BR\left( K^+ \to \pi^+ X \right)_{\rm NA62}< (0.5-2.0)\times 10^{-10}$ at $90\%$ C.L. for the low $m_X$ region~\cite{NA62talks2}. By applying its bound shape (which is related to the $K^+\to\pi^+ X$ efficiency)  to the combined 2016 and 2017 data, we arrive at a slightly better bound,
\begin{eqnarray}
\BR\left(K^+ \to \pi^+ X \right)_{\rm NA62}
< (0.4-1.6)\times 10^{-10},
\label{eq:upperlimitforKptopiX}
\end{eqnarray}
for $m_X\in(0,100)~\MeV$. We will make our further analysis based on this bound. A similar limit, $\BR\left(K^+\to \pi^+ X\right)_{\rm E949} <  (0.5-1.2)\times 10^{-10}$, was obtained by E787/E949~\cite{Artamonov:2009sz}.

\begin{figure}[t!]
\hspace{-0.5cm}
\includegraphics[width=0.45\textwidth]{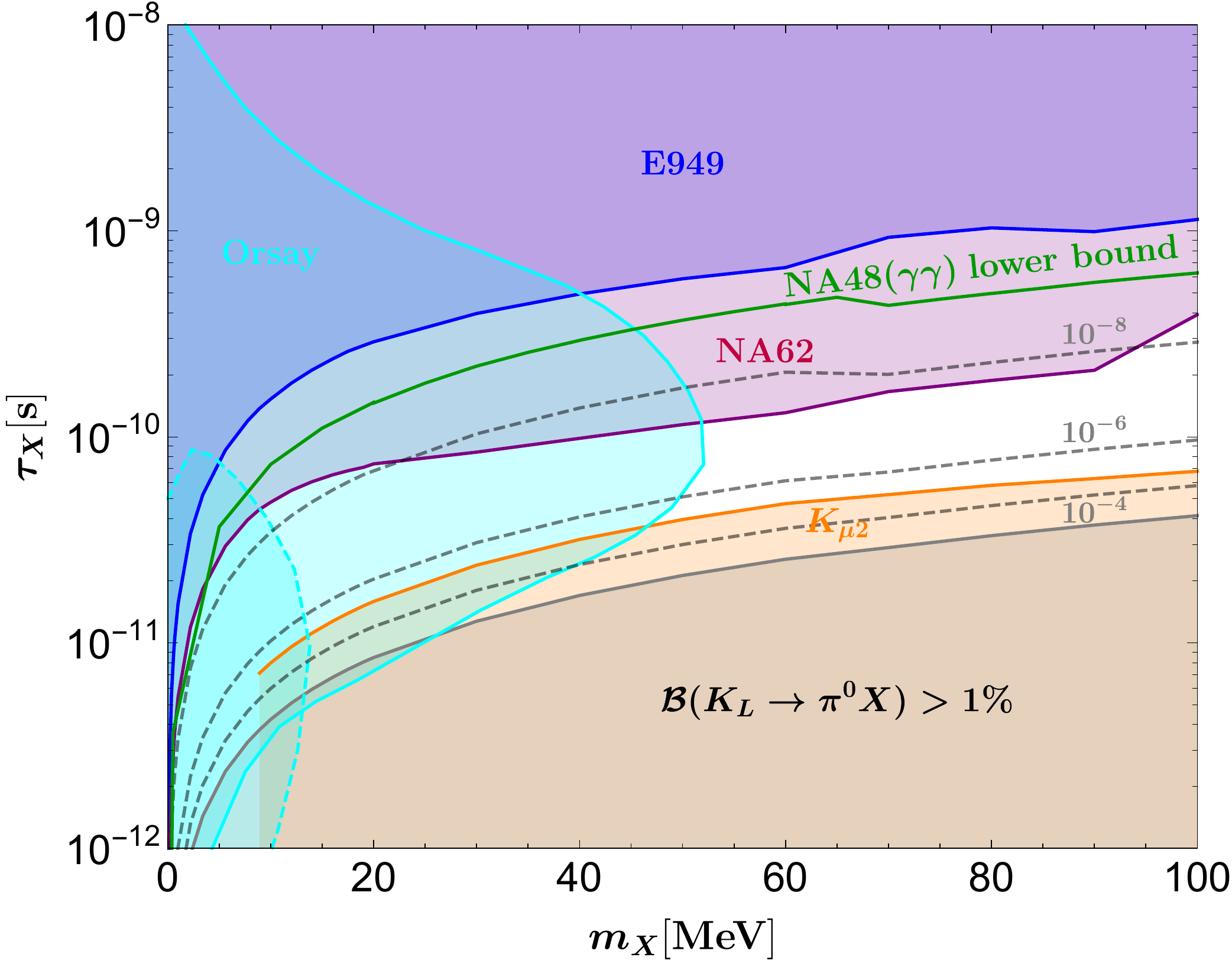}
\caption{Excluded regions in the $m_X - \tau_X$ plane for the decay $K_L \to \pi^0 X$. The purple (blue) region is excluded by the GN bound using measurements of $K_L$ at KOTO~\cite{kototalks} and of $K^+$ at NA62~\cite{NA62talks,NA62talks2} (E949~\cite{Artamonov:2009sz}). The orange region corresponds to the exclusion of $K^+\to\pi^+X$ from $K_{\mu 2}$~\cite{Yamazaki:1984vg}. The gray region is excluded by the untagged $K_L$ decay branching ratio~\cite{Tanabashi:2018oca}. The cyan region bounded by the solid (dashed) curve is excluded by Orsay~\cite{Davier:1989wz} (\cite{Zhitnitsky:1979cn}). The black dashed curves correspond to $\mathcal{B}(K_L \to \pi^0 X) = 10^{-4}$, $ 10^{-6}$ and $10^{-8}$, and the green curve is the lower bound on $\tau_X$ from NA48~\cite{Lai:2002kf} assuming $X\to\gamma\gamma$.}
\label{fig:parameterspace}
\end{figure}

Now we can employ the above results to obtain a strong constraint on the branching ratio of $K_L\to\pi^0X$. We first recall that the real, physical branching ratio may differ its measured value in a specific detector (det) in an experiment such as KOTO or NA62 if $X$ decays~\cite{Kitahara:2019lws}:
\begin{eqnarray}
\BR\left( K \to \pi X \right)_{\rm real} =\BR\left( K \to \pi X\right)_{\rm det} e^{\frac{L}{p}\frac{m_X}{c \tau_X}},
\label{eq:XSbr-real}
\end{eqnarray}
where $\tau_X$ is the lifetime of $X$, and $p/m_X$ and $L$ are the effective boost factor and detector size, respectively. We apply the above relation to both $K_L\to\pi^0X$ at KOTO and $K^+\to\pi^+X$ at NA62, and employ equation~\eqref{eq:GNb} for real branching ratios to link the two, so that we have
\begin{eqnarray}
&&(r-1)\ln\BR(K_L)_{\rm real}
\nonumber
\\
&\gtrsim&r\ln\BR(K_L)_\koto-\ln\left[4.3\BR(K^+)_{\na }\right],
\end{eqnarray}
where $\BR$ with a subscript indicates its measured or real value, and $r=(L/p)_\na/(L/p)_\koto$. We use the values $(L/p)_{\rm KOTO}=(3~\textrm{m})/(1.5~\GeV)$ and $(L/p)_{\rm NA62}=(150~\textrm{m})/(47~\GeV)$. One word of explanation concerning the choice $p\approx 47~\GeV$ at NA62 follows. The $X$ particle is emitted spherically symmetrically from the $K^+$ decay at rest. When the latter is boosted to an energy of $75~\GeV$, we analyzed that the momentum of the $X$ is evenly distributed in the range $p_X\in(40,60)~\GeV$ for the momentum of the $\pi^+$ that falls in the signal region at NA62, $p_{\pi^+}\in(15,35)~\GeV$~\cite{NA62talks2}. Upon comparing with Ref.~\cite{Liu:2020qgx} we found that the choice $p_X\approx 47~\GeV$ fits best their simulation result and is also consistent with the result in Ref.~\cite{Kitahara:2019lws}. We note in passing that the stated choice $E_X\approx 37~\GeV$ in Ref.~\cite{Kitahara:2019lws}, though followed in many later papers, actually neither fits its own numerical result nor falls in the observed $p_{\pi^+}$ range at NA62. Plugging in the central value of $\BR(K_L)_\koto$ from Fig.~\ref{fig:brdistriXS} and the upper bound on $\BR(K^+)_\na$ in equation~\eqref{eq:upperlimitforKptopiX}, we arrive at the lower bound:
\begin{eqnarray}
\BR\left( K_L \to \pi^0 X\right)_{\rm real}> (0.5-3.1)\times 10^{-8},
\label{eq_bound}
\end{eqnarray}
at $90\%$ C.L. for $m_X \in (0, 100)~\MeV$. Applying this bound to equation~\eqref{eq:XSbr-real} for the KOTO detector, we obtain an upper bound on the lifetime $\tau_X$ of $X$ as a function of $m_X$ shown in Fig.~\ref{fig:parameterspace}. A similar analysis is done for the E949 experiment which searched for the two-body decay $K^+\to\pi^+X$ of a stopping $K^+$ and has the parameters $L_{\rm E949}=1.5~\textrm{m}$ and $p_X=\sqrt{\lambda(m_K^2,m_\pi^2,m_X^2)}/(2m_K)$ with the usual triangular function $\lambda(a,b,c)=a^2+b^2+c^2-2ab-2bc-2ca$. The $K_{\mu_2}$ experiment~\cite{Yamazaki:1984vg} set an upper bound on $\BR(K^+\to\pi^+ X)$ independently of whether the $X$ decays or not. We translate it via the GN bound into an upper bound on $\BR(K_L)$, and then apply it to the KOTO detector again through equation~\eqref{eq:XSbr-real}. This then yields a lower bound on $\tau_X$ which is also shown in Fig.~\ref{fig:parameterspace}.

This lower bound of order $10^{-8}$ has been strictly restricted or excluded by other measurements. Being light, $X$ can only decay to $e^+e^-$ and/or $\gamma \gamma$. For $X\to e^+e^-$, we have used the light scalar search result at Orsay $eN\to eN\phi$ with $\phi\to e^+e^-$~\cite{Davier:1989wz,Zhitnitsky:1979cn} to obtain the constraints shown in Fig.~\ref{fig:parameterspace}. (For other potential constraints, see discussions in Ref.~\cite{Liu:2020qgx}.) We see that there is a small survival space for $m_X<60~\GeV$, i.e., with $m_X \in(45,60)~\MeV$ and $\tau_X\in(0.04,0.11)~\textrm{ns}$. For $X\to\gamma\gamma$, the limits mainly come from the measurements of $\BR(K_L\to\pi^0\gamma\gamma)$ and its spectrum at KTeV~\cite{Abouzaid:2008xm} and NA48~\cite{Lai:2002kf}. The results $\BR\left( K_L\to\pi^0\gamma\gamma \right)=(1.29\pm 0.03\pm 0.05)\times 10^{-6}$ at KTeV and $\BR\left(K_L\to\pi^0\gamma\gamma\right)= (1.36\pm0.03\pm0.03)\times 10^{-6}$ at NA48 are consistent with the SM prediction $\sim 1\times 10^{-6}$~\cite{Ecker:1987fm,DAmbrosio:1996kjn,Gabbiani:2002bk,Truong:1993va}. Furthermore, both theoretical calculations and experimental observations give a consistent distribution which is dominated by the invariant mass interval $m_{\gamma\gamma}\in(160,360)~\MeV$. In the following we take the NA48 experiment to describe how to employ its result to set a lower bound on $\tau_X$ for a light $X$ as shown in Fig.~\ref{fig:parameterspace}.

The NA48 experiment detects four photons in the decay $K_L\to\pi^0\gamma_3\gamma_4$ with $\pi^0\to\gamma_1\gamma_2$, assuming that all photons originate from the same $K_L$ decay vertex. For a branching ratio satisfying the bound equation~\eqref{eq_bound}, the $X$ particle from the decay $K_L\to\pi^0X$ will mostly decay in the KOTO detector and thus definitely decay in the NA48 detector considering its parameters $L_{\rm NA48}/p_{\rm NA48}=(100~\textrm{m})/(15~\GeV)$. This means that the decay $K_L\to\pi^0X$ with $X\to\gamma\gamma$ could mimic the $K_L\to\pi^0\gamma\gamma$ signal at NA48. Since the $X$ generically decays at a displaced vertex, the momenta of its daughter photons ($\gamma_3\gamma_4$) will have a larger opening angle ($\theta_{34}$) than they would have if they had originated from the same vertex as the other photon pair ($\gamma_1\gamma_2$) due to momentum conservation. Using the kinematical relation $\cos\theta_{34}=1-m_{34}^2/(2E_3E_4)$, where $E_{3,4}$ and $m_{34}^2$ are respectively the energies and invariant mass squared of the photons $\gamma_3\gamma_4$, we see that a given $m_X$ will appear as a smaller $m_{34}<m_X$ in the NA48 analysis; in other words, the true value of $m_X$ looks as though it is being shifted downwards in the invariant mass analysis of NA48. If the lower limit equation~\eqref{eq_bound} held true for a light $X$, one would expect a peak or at least enhancement around $m_{34}<m_X<60~\GeV$ in the data. However, no such enhancement was observed in the experiment. Actually, NA48 reported no signal events for $m_{34}\in [0,40]~\MeV$, and set the limit $\BR\left(K_L\to \pi^0\gamma\gamma\right)<0.6\times 10^{-8}$ ($90\%$ C.L.) for $m_{34}\in[30,110]~\MeV$~\cite{Lai:2002kf}. Using the data provided in~\cite{Lai:2002kf}, we have estimated the upper bound for $m_{34}\in [0,40]~\MeV$, and found it to be about the same as for $m_{34}\in[30,110]~\MeV$. Taking into account the constraint $y=|p_{K}\cdot(k_{3}-k_{4})|/m_{K}^{2}<0.2$ at NA48, we found by a simple simulation that the acceptance for the two-body decay chain $K_L\to\pi^0X$ with $X\to\gamma\gamma$ is about half of that for the three-body decay, which implies $\BR\left(K_L\to\pi^0X\textrm{ with }X\to\gamma\gamma\right)\lesssim 1.2\times 10^{-8}$. We then use it to obtain a lower bound for $\tau_X$ as shown in Fig.~\ref{fig:parameterspace}. We can see that there is no survival space for $K_L\to\pi^0X$ with $X\to\gamma\gamma$. These are great obstacles to the interpretation of the KOTO result in terms of the decay $K_L\to\pi^0X$, especially when $X$ is a light particle.

\section{$K_L \to \gamma \gamma X$}
\label{sec.3}

\begin{figure*}[t!]
\hspace{-0.5cm}
\includegraphics[width=0.32\textwidth]{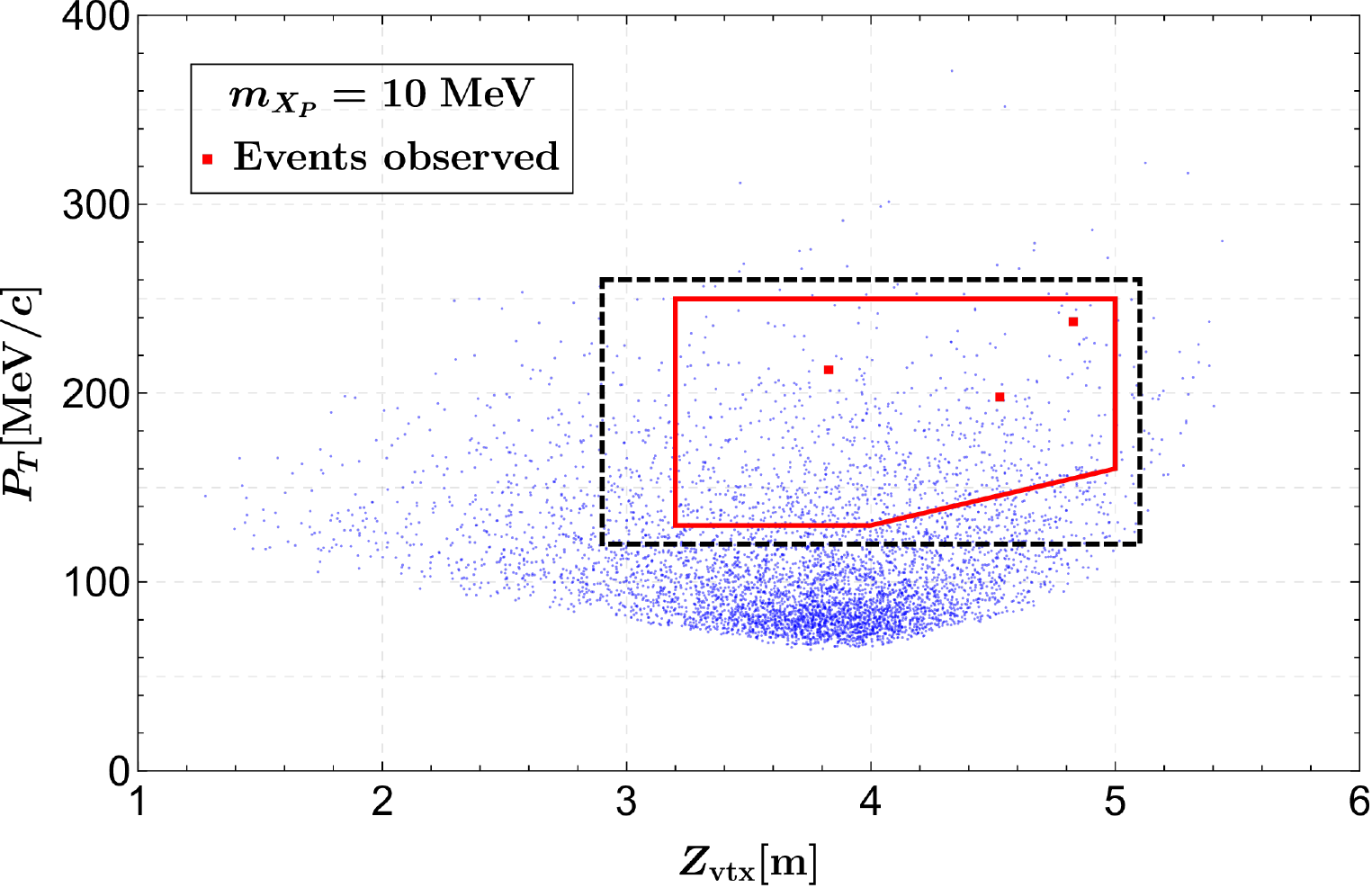} 
\includegraphics[width=0.32\textwidth]{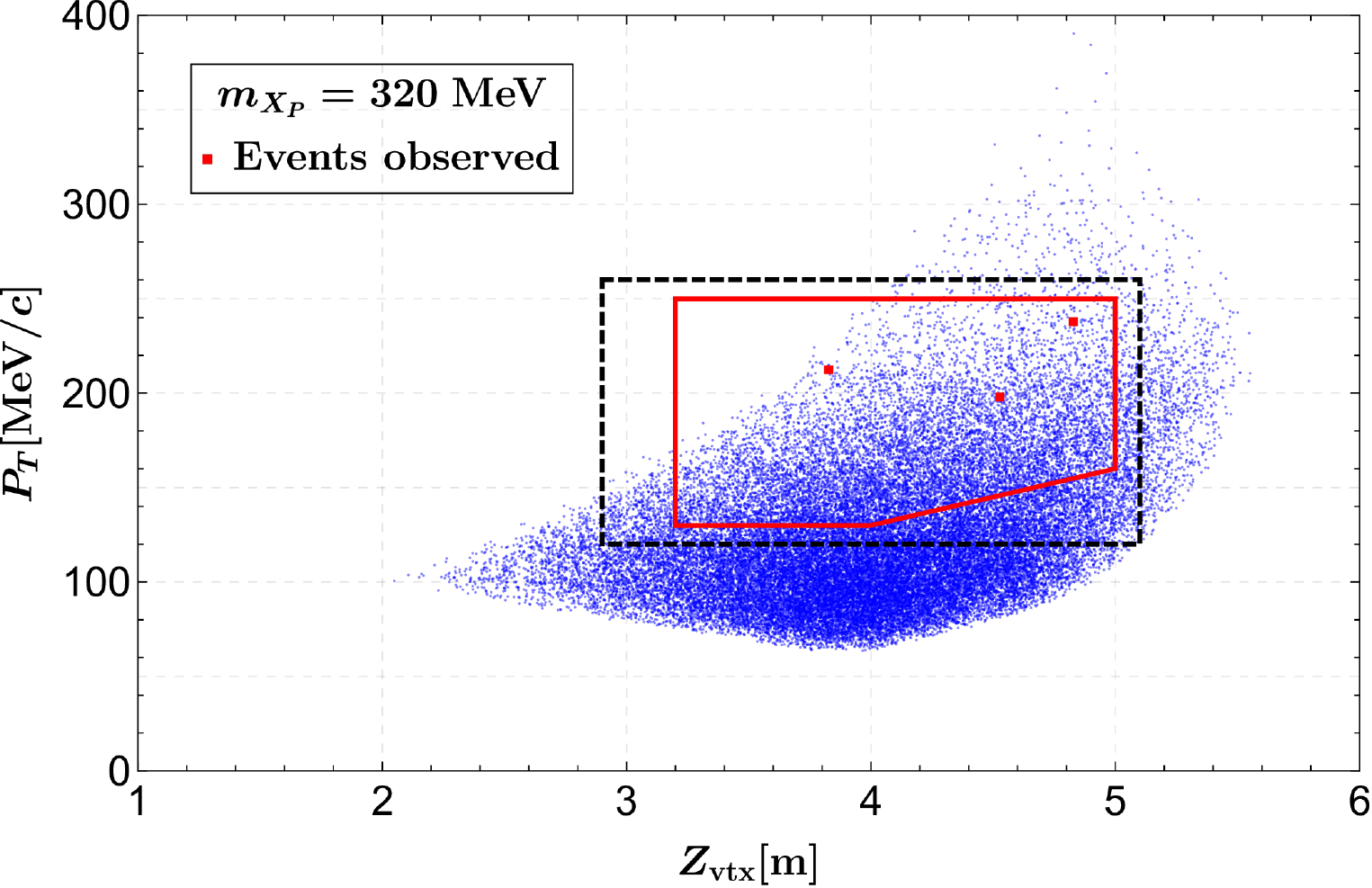} 
\includegraphics[width=0.32\textwidth]{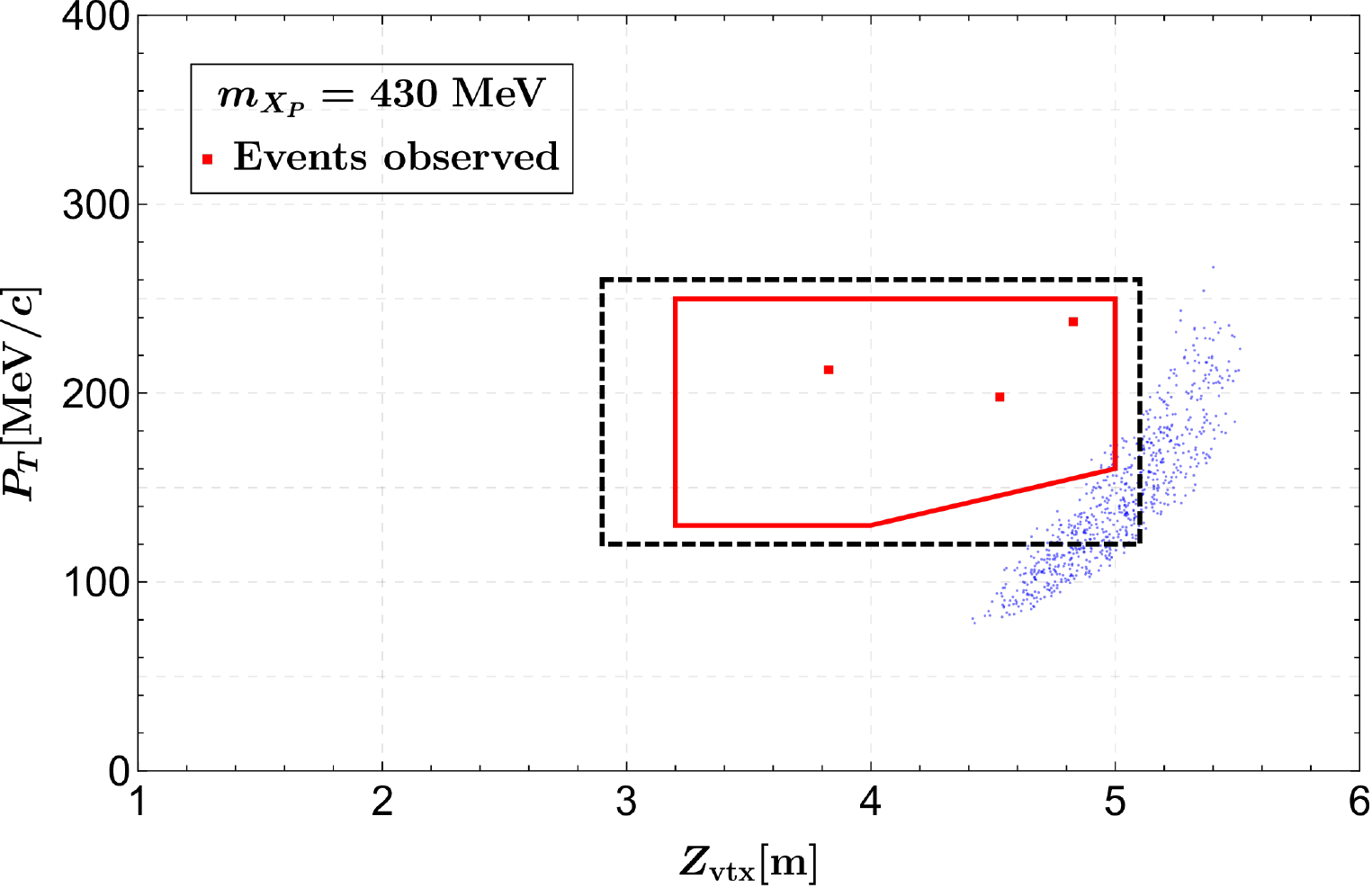} 
\caption{Reconstructed events (blue scatters) in the $Z_\vtx-P_T$ plane after all the cuts for $K_L\to\gamma\gamma X_{P}$ with $m_{X_P} = 10~\MeV$ (left), $320~\MeV$ (middle), and $430~\MeV$ (right). With the same $K_L$ sample the difference in the total number of scatters for various $m_{X_P}$ arises from different cut efficiencies.}
\label{fig:eventdistrXP}
\end{figure*}

In the KOTO experiment the directions of the photon pair were not recorded, and the reconstruction of the decay $K_L\to\pi^0\nu\bar\nu$ was based on the assumption that the photon pair arises from the $\pi^0$ decay which in turn is a product of the $K_L$ decay in the beam line. This motivates us to consider the decay $K_L\to\gamma\gamma X$ in which the photon pair would be misidentified as coming from the $\pi^0$ decay and $X$ appears as missing energy. It is worth mentioning that this process is not constrained by the GN bound as its counterpart in the charged sector is absent. Hence the $X$ here could, in principle, act as dark matter, but this scenario is not supported by the preliminary KOTO result of distributions discussed below.

The decay $K_L\to \gamma\gamma X$ can appear at the one-loop order. The effective field theory calculation of the decay rate is delegated to Appendixes~\ref{Detailedcalc}. Assuming parity symmetry, $X$ could be a pseudoscalar or an axial vector. In this section, we analyze its phenomenological aspects by simulations. The distribution in the $Z_\vtx-P_T$ plane of the reconstructed events is depicted in Fig.~\ref{fig:eventdistrXP} at three typical masses of the pseudoscalar $X_P$. A similar distribution was found for the axial vector case and thus not shown separately. As we can see from the figure, the KOTO's three candidate events could be covered over a range of $m_X$, but the decay events more tend to be located below the signal region. If we naively insist that the KOTO's three events are faked by $K_L\to \gamma\gamma X$, more events should be found below the signal region, which however is not the case. Therefore, we will not pursue this idea further, but employ KOTO's results to work out an experimental upper limit on the decay $K_L\to \gamma\gamma X$ instead.

\begin{figure}[t!]
\hspace{-0.5cm}
\includegraphics[width=0.42\textwidth]{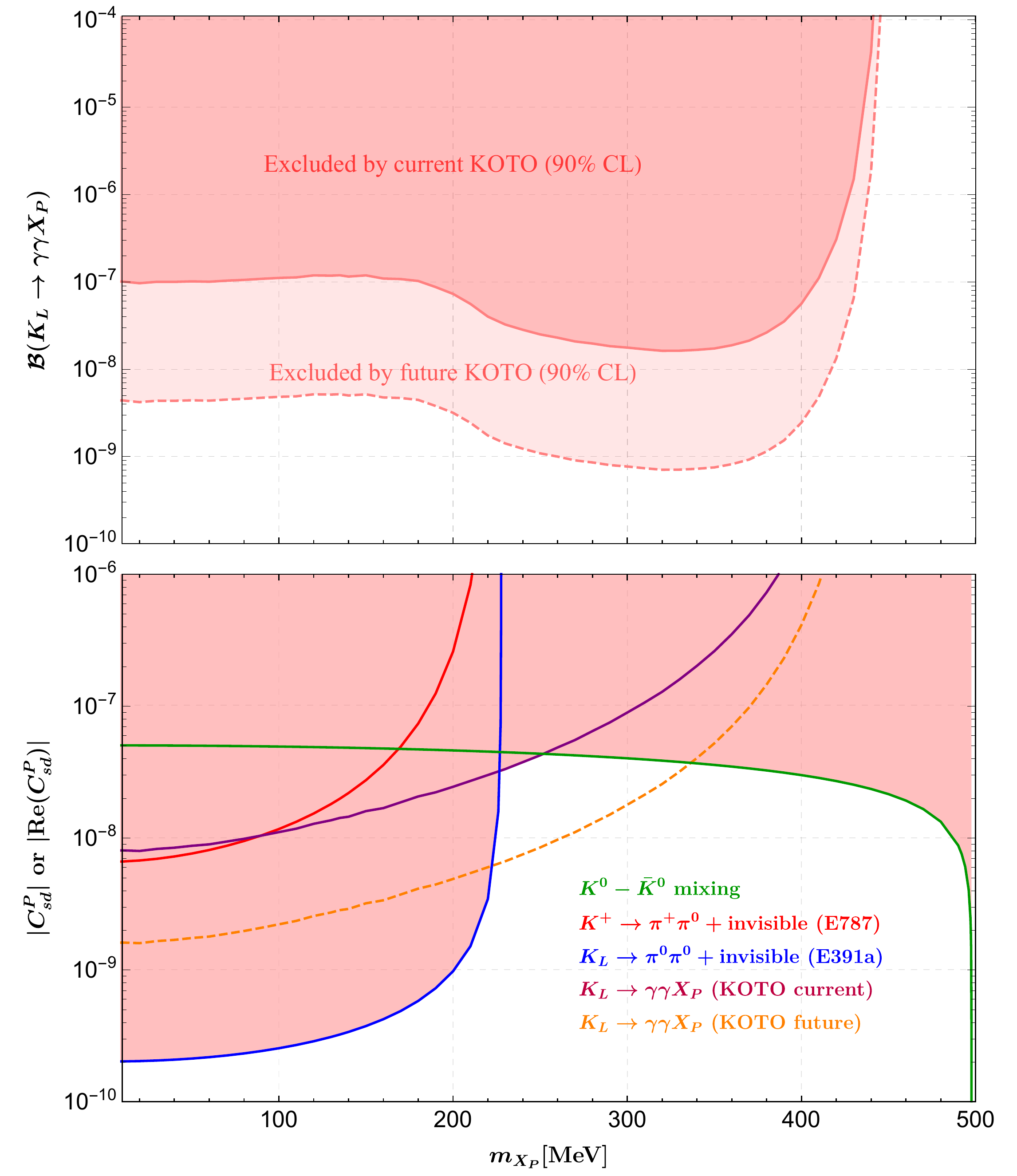}
\caption{Top: upper limit on $\BR(K_L\to\gamma\gamma X_P)$ from KOTO's current data (solid) and projected future capability (dashed). Bottom: upper limits on Wilson coefficients: $|C_{sd}^P|$ from $K^+\to\pi^+\pi^0+\textrm{invisible}$ by E787 experiment (red solid), $|{\rm Re} (C_{sd}^P)|$ from $K^0-\bar{K}^0$ mixing (green solid), $K^+\to\pi^+\pi^0+\textrm{invisible}$ by E391a experiment (blue solid), $K_L\to\gamma\gamma X_P$ by KOTO's current result (purple solid) and future expectation (orange dashed).}
\label{fig:brdistriXP}
\end{figure}

\begin{figure}[t!]
\hspace{-0.5cm}
\includegraphics[width=0.42\textwidth]{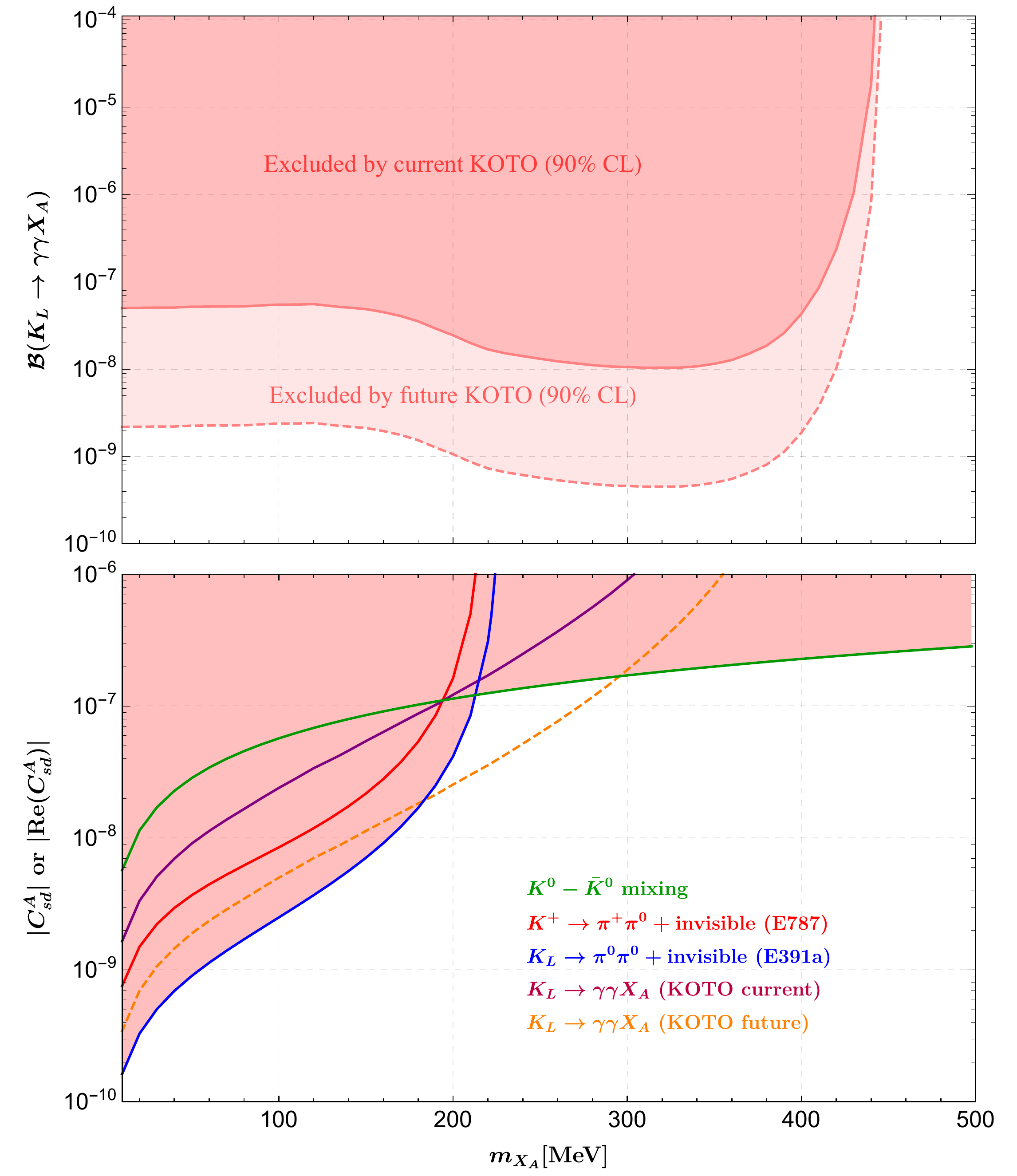}
\caption{Same as in Fig.~\ref{fig:brdistriXP} except for an axial vector $X_A$.}
\label{fig:brdistriXA}
\end{figure}

We choose as our signal region for $K_L\to \gamma\gamma X$ specified by $Z_\vtx\in[2.9,5.1]~{\rm m}$ and $P_T\in[0,120]~\MeV$, which excludes KOTO's three events but includes most of our decay events with a background of $0.08\pm 0.06$. According to the observed zero event in the region~\cite{kototalks}, we obtain an upper limit on $\BR(K_L\to \gamma\gamma X_P)$ and relevant Wilson coefficients as a function of $m_{X_P}$ in the top and bottom panels of Fig.~\ref{fig:brdistriXP}, respectively. Also included is the projected future detection capability of KOTO (dashed curves) when an $\ses_{\pi^0\nu\bar{\nu}} =3.0\times10^{-11}$ is reached. In the bottom panel, we also display limits coming from other experiments: $K^+\to \pi^+\pi^0\nu\bar{\nu}$ at E787~\cite{Adler:2000ic}, $K_L\to\pi^0\pi^0\nu\bar{\nu}$ at E391a~\cite{E391a:2011aa}, and the $K^0-\bar{K}^0$ mixing~\cite{Tanabashi:2018oca}. For a light $X_P$, the limit is dominated by $K_L\to\pi^0\pi^0X_P$, while for a heavy $X_P$ the strongest limit comes from the $K^0-\bar{K}^0$ mixing; in between (roughly for $m_{X_P}\in(220,350)~\MeV$) the KOTO can yield the best upper bound in the future. Similar limits for an axial vector $X_A$ are shown in Fig.~\ref{fig:brdistriXA}.

\section{$K_L \to \pi^0 XX$}
\label{sec.4}

\begin{figure*}[t!]
\hspace{-0.5cm}

\includegraphics[width=0.32\textwidth]{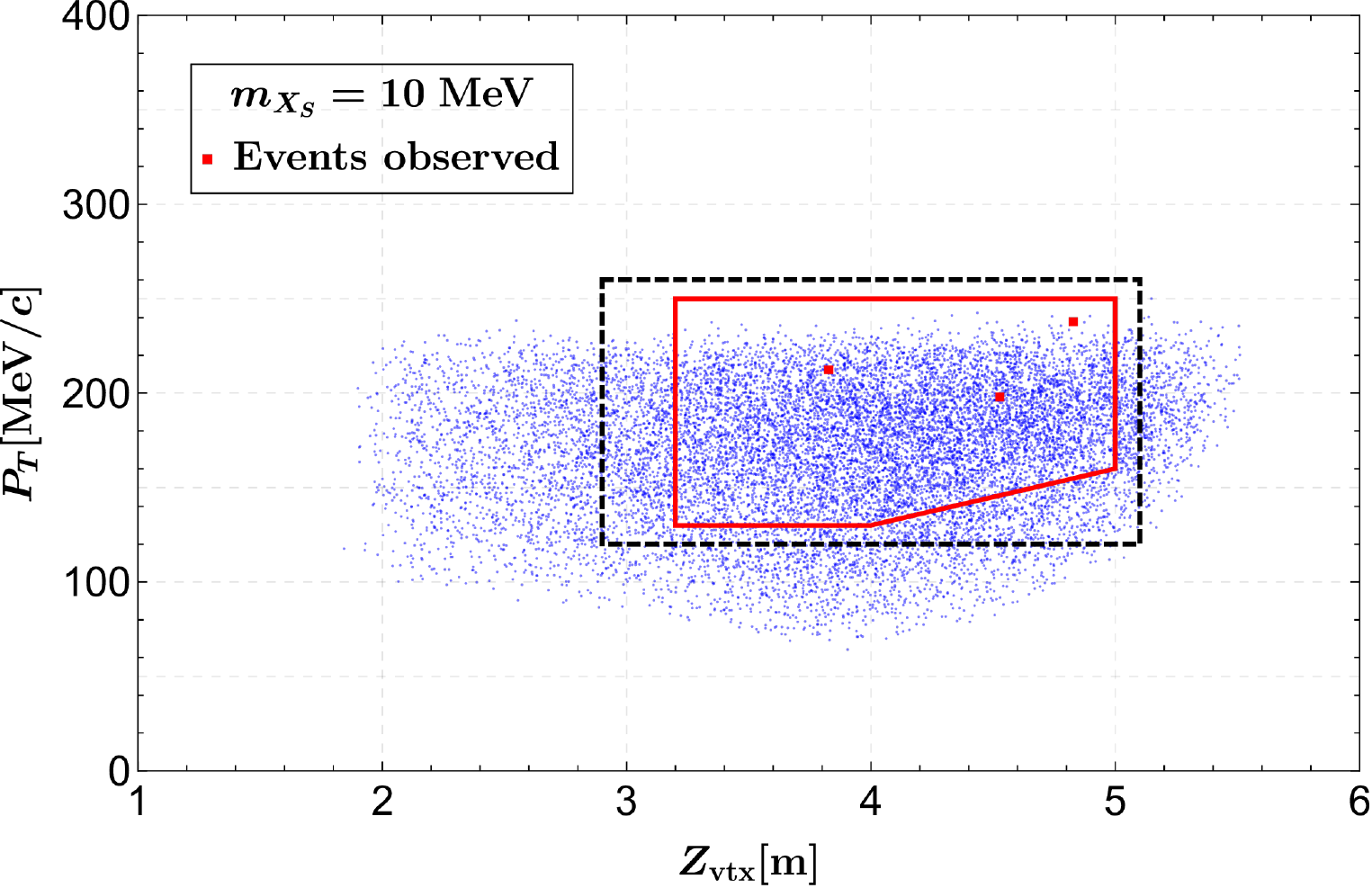} 
\includegraphics[width=0.32\textwidth]{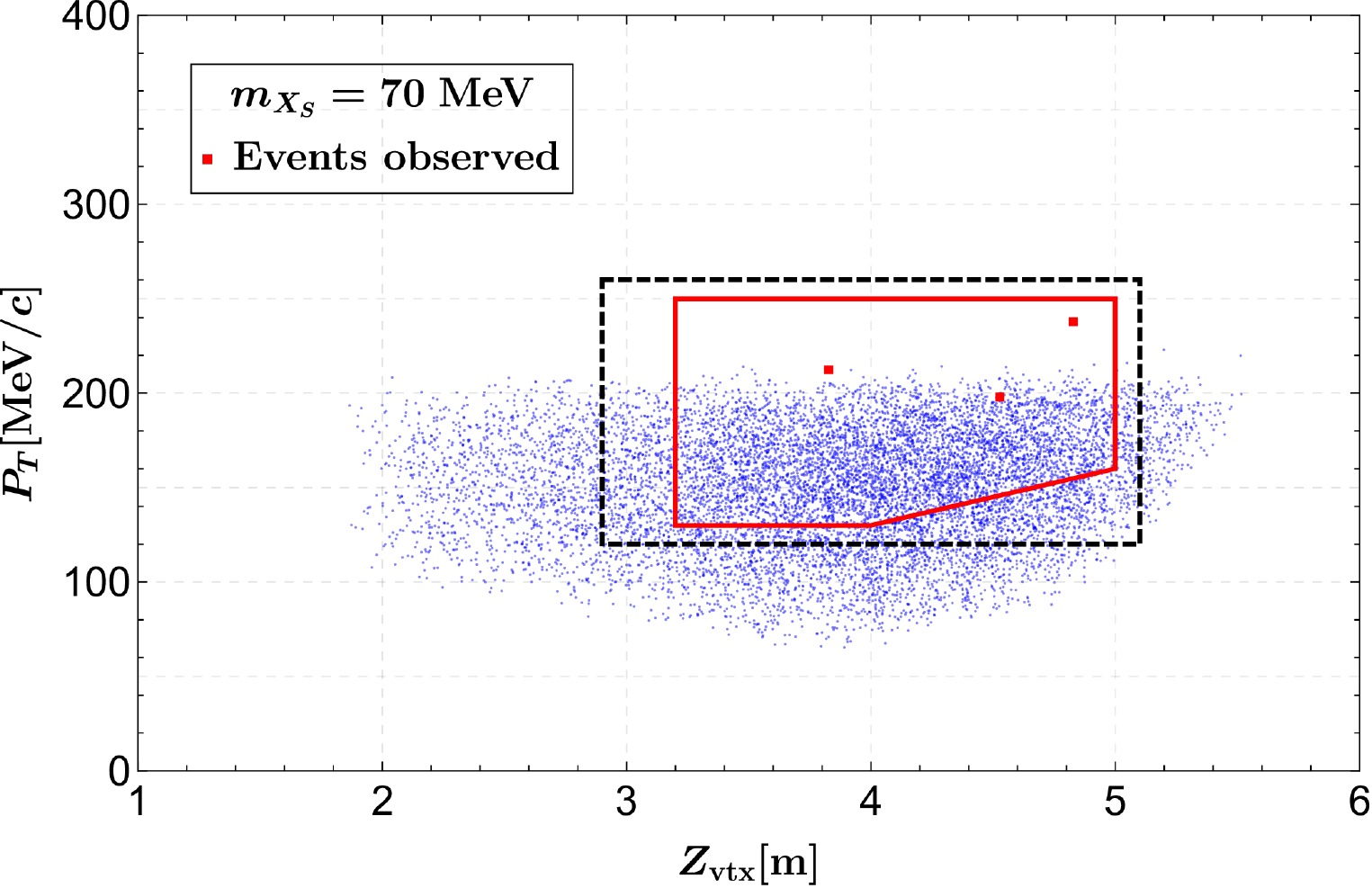} 
\includegraphics[width=0.32\textwidth]{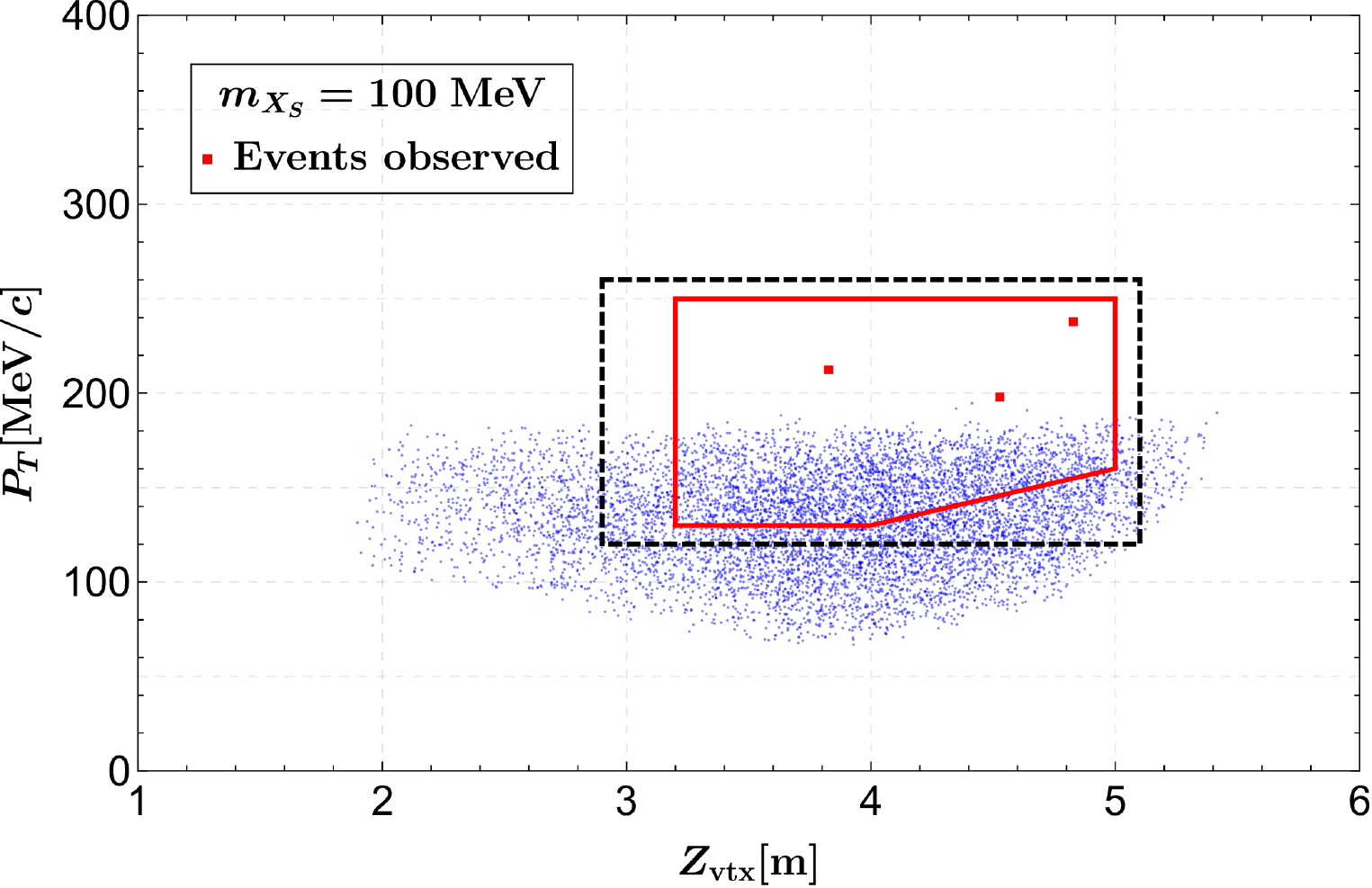} 
\includegraphics[width=0.32\textwidth]{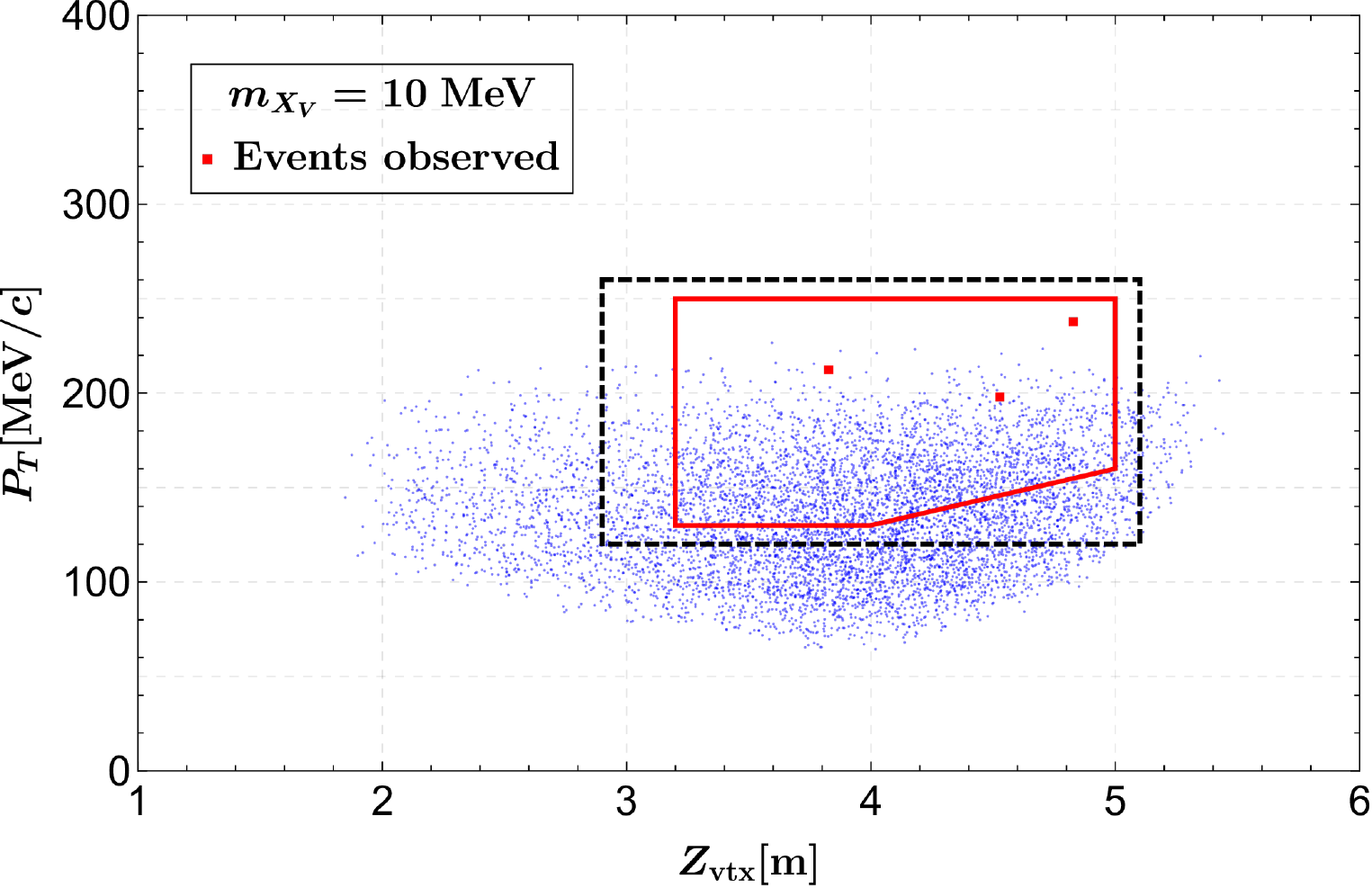} 
\includegraphics[width=0.32\textwidth]{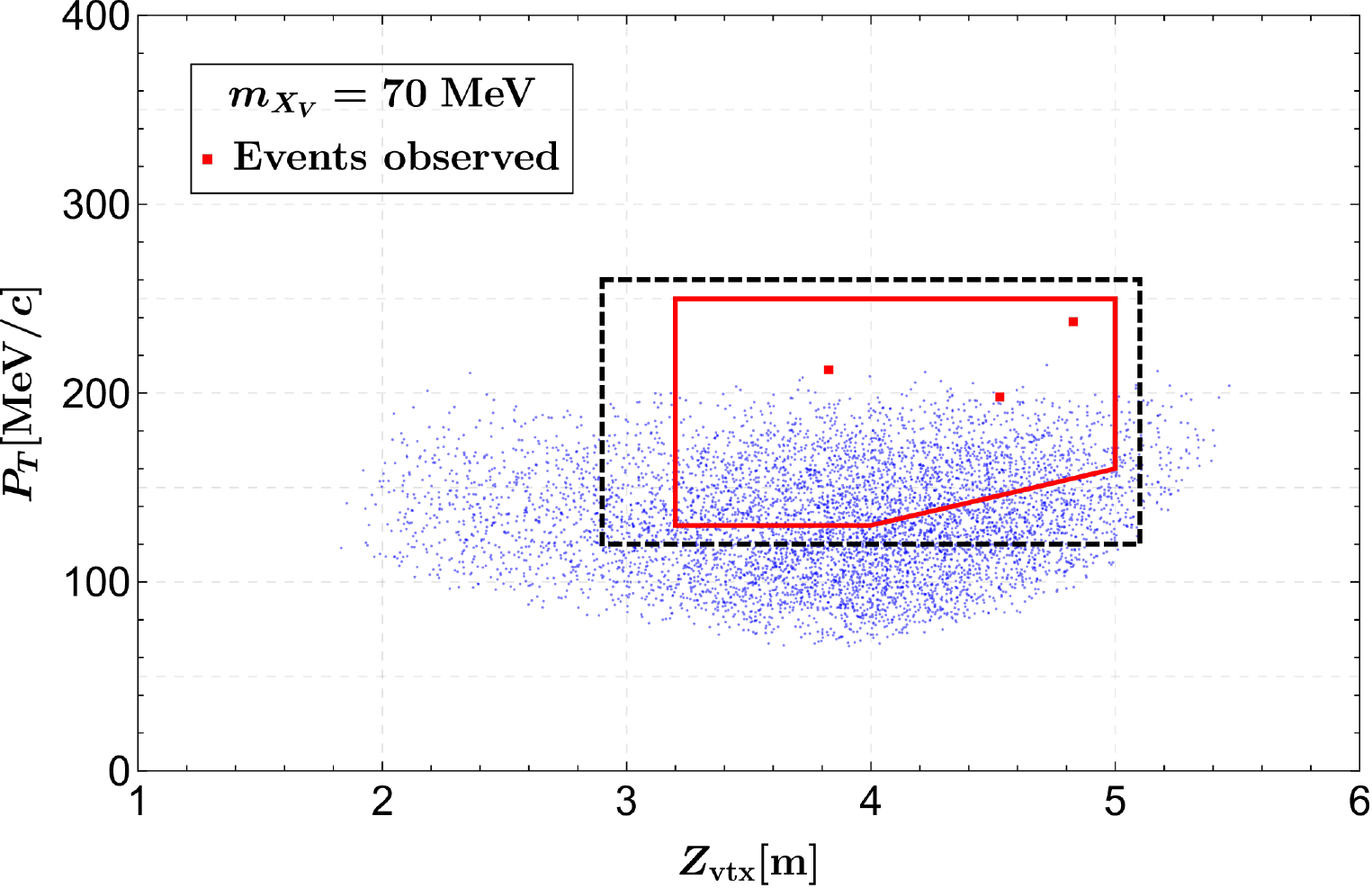} 
\includegraphics[width=0.32\textwidth]{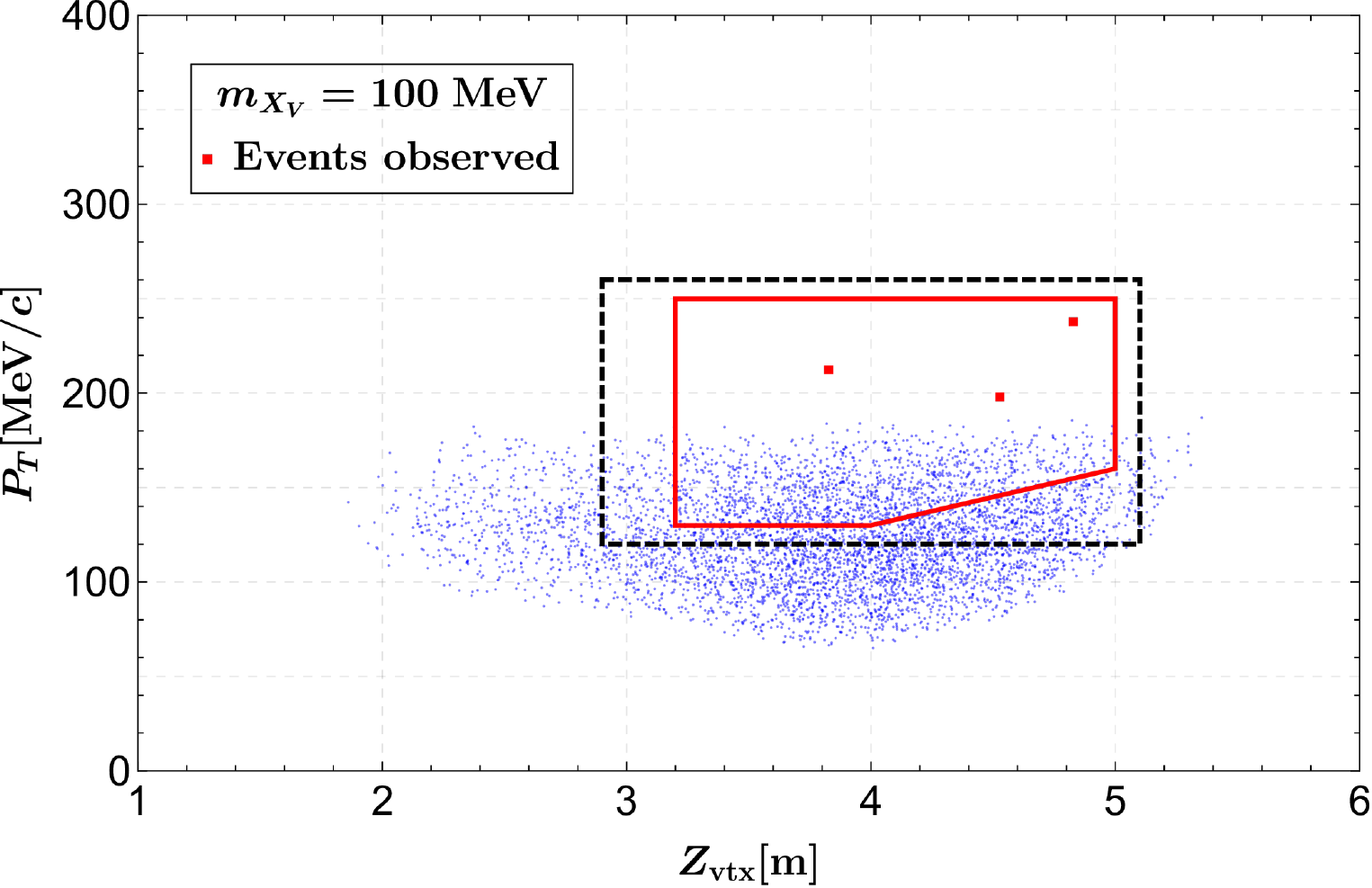} 

\caption{Reconstructed events (blue scatters) in the $Z_\vtx-P_T$ plane after all the cuts for $K_L\to\pi^0XX$ with $m_X=10~\MeV$ (left), $70~\MeV$ (middle), and $100~\MeV$ (right) for a scalar (top) or vector (bottom) $X$.}
\label{fig:eventdistrpixx}
\end{figure*}

\begin{figure}
\hspace{-0.9cm}

\includegraphics[width=0.42\textwidth]{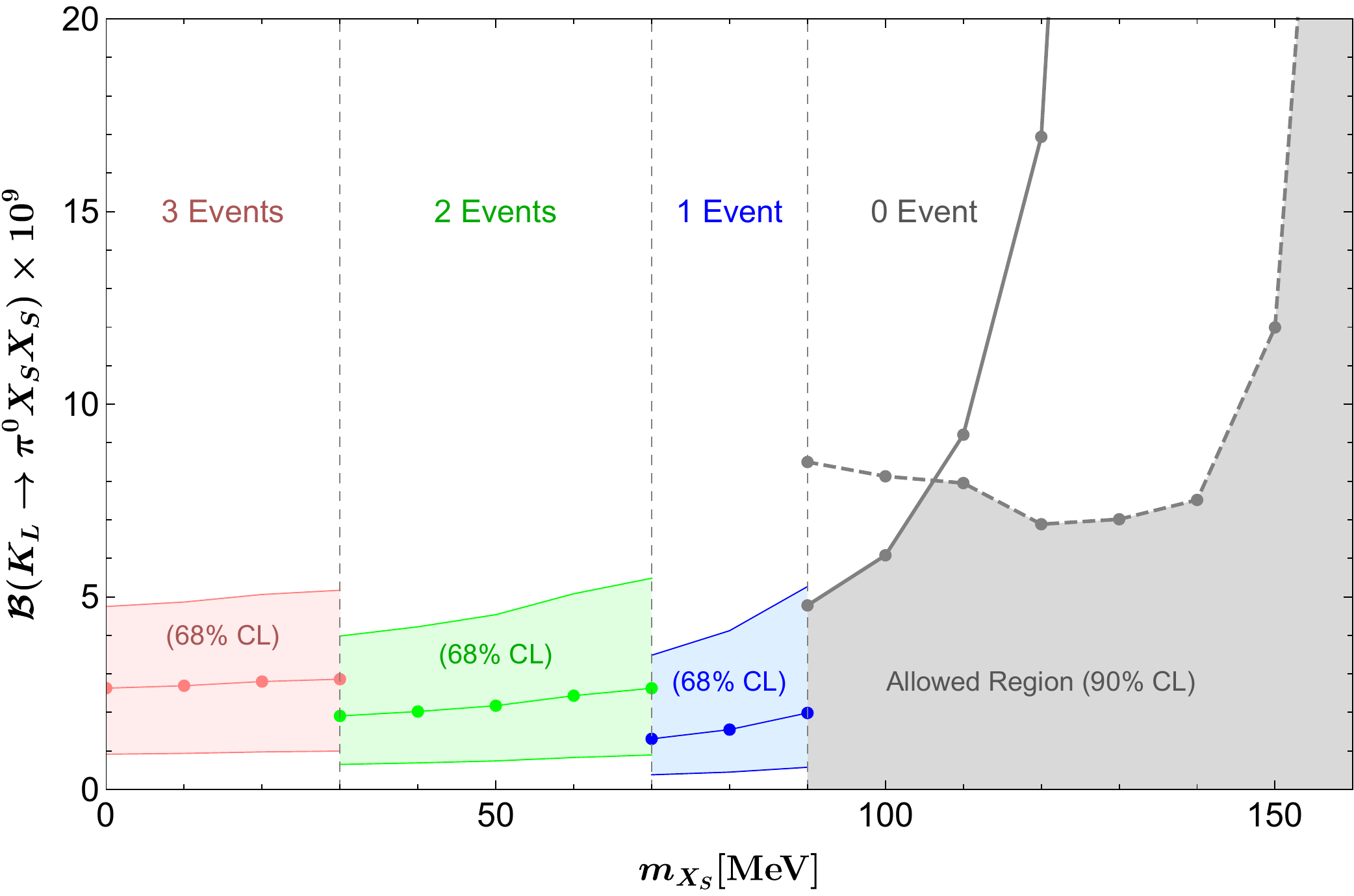} 
\includegraphics[width=0.42\textwidth]{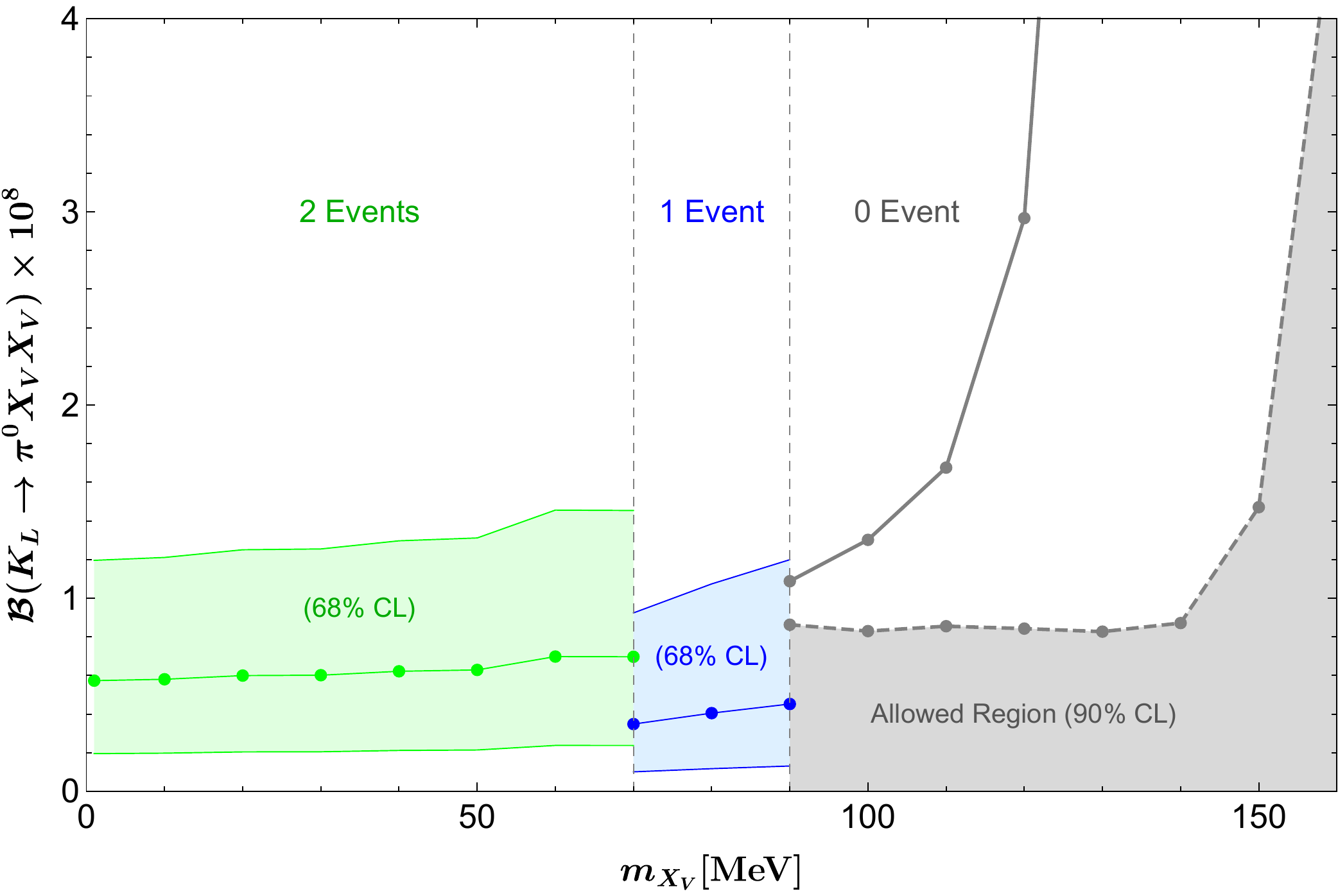} 

\caption{Same as Fig.~\ref{fig:brdistriXS} but for $K_L\to\pi^0XX$ with a scalar (top) or vector (bottom) $X$.}
\label{fig:pixxsvbr}
\hspace{-0.9cm}
\end{figure}

Finally, we discuss the case in which a pair of $X$ particles appear in the $K$ meson decays for which the kinematics will be very different from that of $K\to\pi X$ that we studied in Sec.~\ref{sec.2}. For a fermionic $X$ this has been suggested in Ref.~\cite{Fabbrichesi:2019bmo}. We consider here a bosonic $X$ which, for one reason or another, only couples in pairs to the quarks and may be a scalar or vector of either parity. Our simulation results for the signal distribution in the $Z_\vtx-P_T$ plane are shown in Fig.~\ref{fig:eventdistrpixx} at three different masses. We find that the events for a light $X$ are almost evenly distributed in the signal region as in the case for $K_L\to\pi^0\nu\bar{\nu}$ but differ significantly from the decay $K_L\to\pi^0X$ as shown in Fig.~\ref{fig:eventdistr}. As $X$ becomes heavier, the maximal $P_T$ also drops. But the truncation of the distribution around the maximal $P_T$ is not as sharp as in the two-body decay, and this leaves some space to explain the KOTO signal with a high $P_T$. For even larger masses, the distribution will shift below the blind box.

The branching ratio for $K_L\to\pi^0XX$ from the above simulation is shown in Fig.~\ref{fig:pixxsvbr}. The KOTO candidate signals prefer a light scalar $X_S$ with $m_{X_S}\lesssim 30~\MeV$, which must be unstable and have an appropriate lifetime to avoid the GN bound. The requirement is that with a high probability the two $X_S$s should be both invisible to KOTO while at least one of them decays into SM particles ($\gamma\gamma$ or $e^+e^-$) which can be vetoed at NA62. This implies in passing that the $X$ particle would unlikely play the role of stable dark matter. Qualitatively speaking, in terms of kinematics, the experimental constraints on the $\pi X_SX_S$ mode are looser than those on the $\pi X_S$ mode. But to assess the feasibility of the scenario, we have to determine the appropriate lifetime and mass of $X_S$. The physical branching ratio for $K\to\pi X_SX_S$ at KOTO or NA62 can be expressed as
\begin{eqnarray}
&&\BR(K\to\pi X_SX_S)_{\rm real}
\nonumber
\\
&=&\BR(K\to\pi X_SX_S)_{\rm det} \int
e^{\left(\frac{L_1}{p_1}+\frac{L_2}{p_2}\right)\frac{m_{X_S}}{c \tau_{X_S}}}\times
\nonumber
\\
&&f(L_1,L_2,p_1,p_2)~dL_1dL_2dp_1dp_2,
\label{eq:XXSbr-real}
\end{eqnarray}
where $f$ represents the probability that the two $X_S$'s propagate respectively at the momentum $p_1,~p_2$ for a distance $L_1,~L_2$ to exit the detector without decay. The integration over the signal region ensures that all situations of the $K$ decay have been taken into account. However, it is difficult for us to manage a systematic simulation to cover all information due to the complexity of the displaced-vertex simulation. Considering that the two $X_S$s have a large boost at both KOTO and NA62, they should have a high probability of flying in the beam direction with a small angle to the veto plates. For further estimation, we found that $p_1=p_2\equiv E$ and $L_1=L_2\equiv L$ serve as a good approximation, which leads to the simplification:
\begin{eqnarray}
&&\BR(K\to\pi X_SX_S)_{\rm real}
\nonumber
\\
&\approx&\BR(K\to\pi X_SX_S)_{\rm det}
\int dLdE~e^{\frac{2L}{E}\frac{ m_{X_S}}{c \tau_{X_S}}}F(L,E).
\label{eq:XXSbr-real-approx}
\end{eqnarray}
The two-dimensional probability function $F(L,E)$ is much simpler compared to the four-dimensional one. For $K_L\to\pi^0X_SX_S$ at KOTO, it can be extracted from our current simulation; for $K^+\to\pi^+X_SX_S$ at NA62, we have adopted a rough simulation in which only some cuts on the signal region are considered.

\begin{figure}
\hspace{-0.9cm}
\includegraphics[width=0.42\textwidth]{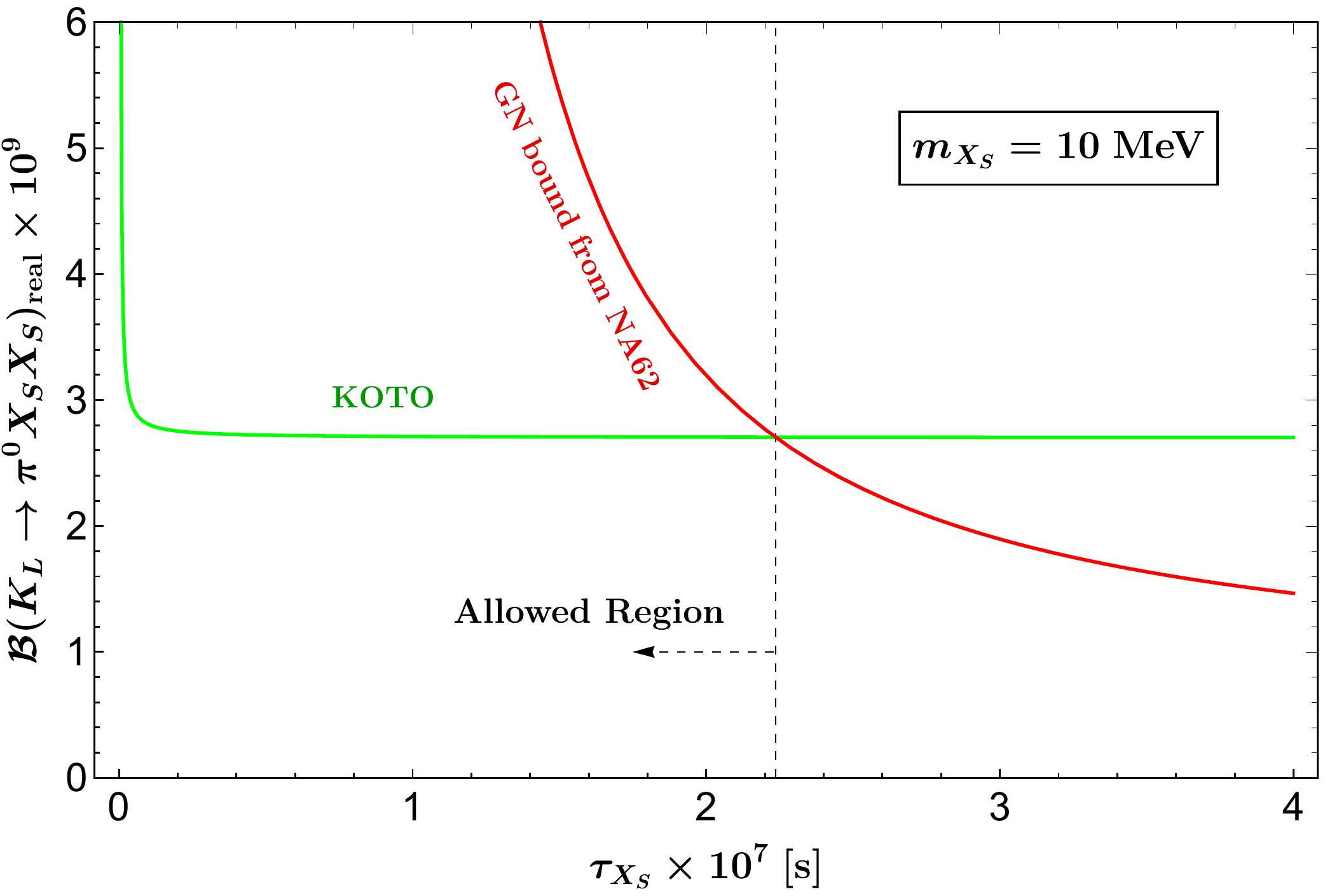} 
\caption{$\BR\left(K_L \to \pi^0 X_SX_S\right)_{\rm real}$ as a function of $\tau_{X_S}$ at $m_{X_S}=10$~MeV (green curve) together with the upper limit from NA62 upon using the GN bound (red).}
\label{fig:brrealKOTONA62}
\hspace{-0.9cm}
\end{figure}

As an example of simulation we consider the scenario that a scalar $X_S$ of mass $m_{X_S}=10~\MeV$ decays into a photon pair. The branching ratio measured at KOTO can be read off in Fig.~\ref{fig:pixxsvbr}, with the central value being $\BR(K_L\to\pi^0 X_SX_S)_{\rm KOTO} =2.7\times 10^{-9}$. For NA62, assuming that $K^+ \to \pi^+X_SX_S$ has the same acceptance as $K^+ \to \pi^+ \nu \bar{\nu}$, we obtain $\BR(K^+\to\pi^+X_SX_S)_{\rm NA62}<1.60\times 10^{-10}$ at 95\%~C.L.. By incorporating all these into equation~\eqref{eq:XXSbr-real-approx}, we obtain $\BR(K_L\to\pi^0X_SX_S)_{\rm real}$ as a function of the lifetime $\tau_{X_S}$ shown in Fig.~\ref{fig:brrealKOTONA62} as the green curve, and an upper limit on it (red curve) from NA62 with the aid of the GN bound. This yields the allowed region with $\tau_{X_S}\lesssim 2.2\times 10^{-7}~\text{s}$, within which the veto information can also be gained from the figure: at least 74.6\% of the $K^+\to\pi^+X_SX_S$ signals are vetoed at NA62, while for $\tau_{X_S}\gtrsim 10^{-8}~\text{s}$ the two $X_S$s at KOTO nearly do not decay before exiting the detector. On the other hand, the untagged $K_L$ branching ratio~\cite{Tanabashi:2018oca} constrains $\BR(K_L \to \pi^0 X_SX_S) < 1\%$, which leads to $\tau_{X_S}\gtrsim 1.7\times 10^{-10}~\text{s}$. But we emphasize once again that this is only a rough estimate, and an appropriate determination of the lifetime and mass of $X_S$ could only be achieved by a systematic detector simulation. We advocate that the KOTO and NA62 Collaborations will take this endeavor in their future experimental analysis.

\section{Conclusion}
\label{sec.5}

We have investigated by simulations the feasibility to interpret the recent KOTO result in terms of a new neutral particle that appears in the kaon decays. Since $\pi^0$  decays into a pair of photons and neutrinos appear as missing energy, we have considered three scenarios, i.e., $K_L\to\pi^0X,~\gamma\gamma X,~\pi^0XX$. Our results can be summarized as follows. The simplest scenario $K_L\to\pi^0X$ is difficult to accommodate all three candidate events at KOTO, especially the one with a high $P_T$. The signal events for a relatively heavy $X$ are mainly distributed below the KOTO's signal region, which have been employed to work out a bound on the decay branching ratio. While the KOTO result tends to favor a light $X$, our comprehensive analysis on the NA62 and other experiments sets a strong constraint on the scenario $K_L\to\pi^0X$: the decay $X\to\gamma\gamma$ is essentially excluded while the decay $X\to e^+e^-$ leaves open a very small region in the $m_X$-$\tau_X$ plane. Since the scenario $K_L\to\gamma\gamma X$ has no constraint at NA62, it is free of the GN bound. While the three candidate events can be accommodated, the distributions do not fit: more events would be expected below the blind box. We have used the latter to set a constraint on the branching ratio, and compared it with those from other measurements and the expected KOTO's future capability. In terms of the signal distribution, the KOTO candidate events favor the third scenario $K_L\to\pi^0X_SX_S$, where $X_S$ is a scalar with $m_{X_S}\lesssim 30~\MeV$. To accommodate the measurements at NA62, $X_S$ should be unstable but long lived, whose lifetime is estimated to be $1.7\times 10^{-10} \ {\rm s}\lesssim \tau_{X_S} \lesssim 2.2\times 10^{-7} \ {\rm s}$ at $m_{X_S}= 10~\MeV$. But a more precise result would necessitate sophisticated simulations which we hope the KOTO and NA62 Collaborations will perform in their future experimental analysis.

\section*{ACKNOWLEDGMENTS}

We would like to thank Hai-Bo Li for helpful discussions, K. Tobioka for electronic communications, and J. Liu for explanations concerning Ref.~\cite{Liu:2020qgx}. This work was supported in part by Grants No. NSFC-11975130, No. NSFC-11575089, the China Postdoctoral Science Foundation Grant No.~2018M641621, by the National Key Research and Development Program of China under Grant No.~2017YFA0402200, and by the CAS Center for Excellence in Particle Physics (CCEPP).

We thank the anonymous referee for informing us of the NA62 group’s preliminary result based on their 2017 data after we submitted the first version of this work.

\appendix

\section{EFFECTIVE FIELD THEORY FRAMEWORK}
\label{EFT}

For our purpose of accounting for the KOTO anomaly we assume a new real neutral particle $X$ of mass below a few hundred MeV. It may be a scalar ($X_S$), pseudoscalar ($X_P$), vector ($X_V$) or axial vector ($X_A$) particle as appropriate to the scenario under consideration. We start with the low energy effective field theory that contains the $X$ particle in addition to the light quarks and leptons and has the QCD and QED gauge symmetries. As we are mainly concerned with the transitions between the down and strange quarks, we only consider their couplings to the $X$ field:
\begin{eqnarray}
{\cL}_X&=&C^{S}_{pr}(\overline{d^p}d^r)X_S
+C^{P}_{pr}(\overline{d^p}i\gamma_5d^r)X_P
\nonumber
\\
&&+C^V_{pr}(\overline{d^p}\gamma_\mu d^r)X_V^\mu+C^A_{pr}(\overline{d^p}\gamma_\mu\gamma_5 d^r)X_A^\mu,
\label{eq_1X}
\end{eqnarray}
where $p,~r$ refer to the $d,~s$ quarks and the Wilson coefficients are $2\times 2$ Hermitian matrices in the $d,~s$ space. When studying the third scenario $K\to\pi XX$, we switch off the above single-$X$ couplings but switch on the following double-$X$ couplings:
\begin{eqnarray}
{\cL}_{XX}=(\bar sd)\left(g_X^SXX+g_X^VX^\mu X_\mu\right)
+\textrm{H.c.},
\label{eq_2X}
\end{eqnarray}
where $X$ and $X_\mu$ may have any parity without affecting our discussions in the work.

Below the chiral symmetry breaking scale the Nambu-Goldstone bosons become the dynamical degrees of freedom, whose low energy physics is determined at leading order by~\cite{Gasser:1983yg,Gasser:1984gg}
\begin{eqnarray}
\mathcal{L}_\chi&=&
\frac{F_0^2}{4}{\rm Tr}\left(D_\mu U (D^\mu U)^\dagger \right)
\nonumber
\\
&&+\frac{F_0^2}{4}{\rm Tr} \left(\chi U^\dagger +U\chi^\dagger \right),
\label{Lchi}
\end{eqnarray}
where $U=\exp\left(i\sqrt{2}\Phi/F_0\right)$ exponentiates the octet Nambu-Goldstone bosons
\begin{align}
\Phi&=\begin{pmatrix}
\frac{\pi^0}{\sqrt{2}}+\frac{\eta}{\sqrt{6}} & \pi^+ & K^+
\\
\pi^- & -\frac{\pi^0}{\sqrt{2}}+\frac{\eta}{\sqrt{6}} & K^0
\\
K^- & \bar{K}^0 & -\sqrt{\frac{2}{3}}\eta
\end{pmatrix},
\end{align}
and
\begin{eqnarray}
&&D_\mu U=\partial_\mu U-i l_\mu U+i U r_\mu,
\\
&&\chi=2B(s-ip).
\end{eqnarray}
As usual, $F_0$ is the decay constant in the chiral limit and $B$ parametrizes the quark condensate $\langle\bar qq\rangle=-3BF_0^2$. The new field $X$ gets involved in the form of additional terms in the external sources to QCD, which are Hermitian matrices in the light flavor space:
\begin{eqnarray}
l_\mu&=&C^VX_V^\mu-C^AX_A^\mu,
\\
r_\mu&=&C^VX_V^\mu+C^AX_A^\mu,
\\
s&=&C^SX_S,
\\
p&=&-C^PX_P,
\end{eqnarray}
in the case of single-$X$ couplings in equation~\eqref{eq_1X}, and
\begin{eqnarray}
s_{pr}&=&\delta_p^3\delta_r^2\left(g^S_XXX+g^V_X X_\mu X^\mu\right)
\nonumber
\\
&&+\delta_p^2\delta_r^3\left(g^{S*}_XXX+g^{V*}_X X_\mu X^\mu\right),
\end{eqnarray}
in the case of double-$X$ couplings in equation~\eqref{eq_2X}. Note that the QED interaction is contained as usual in the sources $l_\mu,~r_\mu$, which we do not write explicitly. The effective interactions contained in equation~\eqref{Lchi} will be applied to calculate the quantities in the following appendices.

\section{DECAY AMPLITUDES AND DISTRIBUTIONS}
\label{Detailedcalc}

The decay width for $K_L\to\pi^0 X$ is, for a scalar $X_S$,
\begin{eqnarray}
\frac{d\Gamma_{\pi^0X_S}}{dp_{\pi,z}} &=&
\frac{B^2}{16\pi m_{K_L}^2}\left[{\rm Re}~C^{S}_{sd}\right]^2,
\label{eq:distpixs}
\end{eqnarray}
where $p_{\pi,z}$ is the $\pi^0$ momentum component in the $K_L$ beam direction, or, for a vector $X_V$,
\begin{eqnarray}
  \frac{d\Gamma_{\pi^0X_V}}{dp_{\pi,z}} &=&\frac{\lambda(m_{K_L}^2,m_{\pi^0}^2,m_X^2)}{16\pi m_{K_L}^2m_X^2}\left[{\rm Im}~C^{V}_{sd}\right]^2,
  \label{eq:distpixv}
\end{eqnarray}
with $\lambda(x,y,z)=x^2+y^2+z^2-2xy-2yz-2zx$.

\begin{figure}[t!]
\hspace{-0.5cm}
\includegraphics[width=0.42\textwidth]{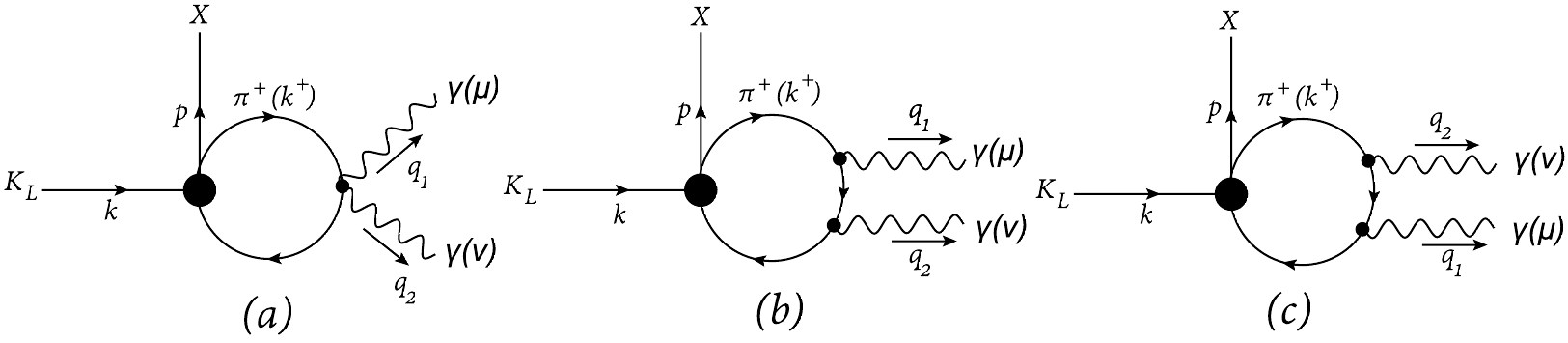} 
\caption{Feynman diagrams for $K_L\to \gamma\gamma X$ at first nonvanishing order. The large (small) dot stands for the effective $X$ (standard QED) interactions in equation~\eqref{Lchi}.}
\label{feykxgg}
\end{figure}

The decay $K_L(k)\to \gamma(q_1)\gamma(q_2)X(p)$ involves only neutral particles, and can only take place at the one-loop order whose Feynman diagrams are shown in Fig.~\ref{feykxgg}. The result is finite, and the (spin-summed) squared matrix element is, for a pseudoscalar $X_P$,
\begin{eqnarray}
|\mathcal{M}_P|^2&=& \left(\frac{\alpha B{\rm Re}~C^{P}_{sd}}{6\sqrt{2} \pi F_0s}\right)^2\left|f(r_\pi)+f(r_K)\right|^2,
\end{eqnarray}
where $r_{\pi,K}=s/(4m_{\pi^\pm,K^\pm}^2)$, $s=(q_1+q_2)^2$, and $\alpha\approx 1/137$ is the fine structure constant, or for an axial vector $X_A$,
\begin{eqnarray}
|\mathcal{M}_A|^2&=& \left(\frac{\alpha{\rm Re}~C^{A}_{sd}}{8\sqrt{2} \pi F_0 m_{X_A}s}\right)^2\left|f(r_\pi)+f(r_K)\right|^2
\nonumber
\\
&&
\times\lambda(m_{K_L}^2, m_{X_A}^2, s).
\end{eqnarray}
The one-loop function is
\begin{equation}
f(r)=4+\left\{
\begin{array}{ll}
\displaystyle -\frac{4}{r}\arcsin^2(\sqrt r),&\!\! r\leq 1
\\
\displaystyle\frac{1}{r}\left[2\ln\left(\sqrt{r}-\sqrt{r-1}\right)
+i\pi\right]^2, &\!\!  r>1
\end{array} \right.
\end{equation}
For simulations, we use the distribution
\begin{eqnarray}
\frac{d\Gamma_{\gamma\gamma X_{P(A)}}}{ds~dt} &=&\frac{1}{512\pi^3 m_{K_L}^3}|\mathcal{M}_{P(A)}|^2,
\label{eq:distpixxg}
\end{eqnarray}
where $t=(k-q_1)^2$.

The decay width for $K_L(k)\rightarrow\pi^0(p) X(q_1) X(q_2)$ is easily computed to be, for a scalar $X_S$ or vector $X_V$ respectively,
\begin{eqnarray}
  \frac{d\Gamma_{\pi^0X_SX_S}}{ds~dt} &=&
  \frac{B^2({\rm Re}~g_X^S)^2}{128\pi^3 m_{K_L}^3},
  \label{eq:distpixxs}
  \\
  \frac{d\Gamma_{\pi^0X_VX_V}}{ds~dt} &=&
  \frac{B^2({\rm Re}~g_X^V)^2}{128\pi^3 m_{K_L}^3}
\left[2+(2r_X-1)^2\right], \label{eq:distpixxv}
\end{eqnarray}
where $r_X=s/(4m_X^2)$, $s=(q_1+q_2)^2$, and $t=(k-q_1)^2$.

\section{OTHER EXPERIMENTAL CONSTRAINTS}
\label{Relevant results}

In this appendix we list other experimental constraints that have been used in Sec.~\ref{sec.3} for a comprehensive analysis. The experimental upper limits on the four-body kaon decays are~\cite{Adler:2000ic,E391a:2011aa}
\begin{eqnarray}
\mathcal{B}(K^+\to \pi^0\pi^+\nu\bar\nu)&<&4.3\times10^{-15},
\\
\mathcal{B}(K_L\to \pi^0\pi^0\nu\bar\nu)&<&8.1\times10^{-7}.
\end{eqnarray}
The corresponding amplitudes for the decay $K(k)\rightarrow\pi(p_1)\pi(p_2) X(q)$ read, for a pseudoscalar $X_P$, \begin{eqnarray}
\mathcal{M}(K^+\rightarrow \pi^+\pi^0X_P)&=&\frac{B}{2F_0}C^{P}_{sd}\frac{u-t}{m_K^2-m_X^2},
\\
\mathcal{M}(\bar{K}^0\rightarrow \pi^0\pi^0X_P)&=&
\frac{BC^{P}_{ds}}{2\sqrt2F_0}\frac{s+m_X^2-m_K^2}{m_K^2-m_X^2},
\end{eqnarray}
and for an axial vector $X_A$,
\begin{eqnarray}
\mathcal{M}(K^+\rightarrow \pi^+\pi^0X_A)&=&
i\frac{C^A_{sd}}{F_0}(p_2-p_1)\cdot \epsilon^*,
\\
\mathcal{M}(\bar{K}^0\rightarrow \pi^0\pi^0X_A)&=&
i\frac{C^A_{ds}}{\sqrt 2F_0}k\cdot \epsilon^*,
\end{eqnarray}
where $s=(p_1+p_2)^2$, $t=(k-p_1)^2$, $u=(k-p_2)^2$, and $\epsilon$ is the polarization vector of $X_A$.

Moreover, the $K^0-\bar K^0$ mixing can give limits on the couplings. The experimental measured quantities are the $K_L-K_S$ mass difference $\Delta M_K$ and the CP violation parameter $\epsilon_K$~\cite{Buras:2005xt,Nierste:2009wg,He:2005we}, whose current experimental values are~\cite{Tanabashi:2018oca},
\begin{eqnarray}
\Delta M_K&=&(3.484\pm 0.006)\times10^{-12}~{\rm {MeV}},
\\
|\epsilon_K|&=&(2.228\pm 0.011)\times10^{-3}.
\label{eq:epex}
\end{eqnarray}
Considering the theoretical uncertainties from long-distance contributions in $\Delta M_K$, we require the new contribution do not exceed the experimental value. Relatively, the calculation of $\epsilon_K$ is more credible; hence we, require the new contribution to be less than 30\% of its experimental value. The limits on the Wilson coefficients then read, for a pseudoscalar $X_P$,
\begin{eqnarray}
&&\left|({\rm Re}~C^{P}_{sd})^2-({\rm Im}~C^{P}_{sd})^2\right|
<\frac{1}{2}R_A,
\\
&&\left|{\rm Re}~C^{P}_{sd}~{\rm Im}~C^{P}_{sd}\right|
<\frac{0.3}{\sqrt2}|\epsilon_K|R_A,
\end{eqnarray}
with $R_P=\Delta M_K m_K(m_K^2-m_{X_P}^2)/(B^2F_K^2)$, and for an axial vector $X_A$,
\begin{eqnarray}
&&\left|({\rm Re}~C^A_{sd})^2-({\rm Im}~C^A_{sd})^2\right|<\frac{1}{2}R_A,
\\
&&\left|{\rm Re}~C^A_{sd}~{\rm Im}~C^A_{sd}\right|<\frac{0.3}{\sqrt2}|\epsilon_K|R_A,
\end{eqnarray}
with $R_A=\Delta M_K m_{X_A}^2/(F_K^2 m_K)$.

\section{SIMULATION OF $K_L$ DECAY}
\label{Simulation}

In this appendix we will briefly describe how we do the simulations for various $K_L$ decays. A systematic simulation is complicated and time consuming, so we content ourselves with a simplified version of it in this work. We find that even in this simple framework we can get accurate results as in a systematic simulation. In the following, we will explain our procedure and compare our result on $K_L\to\pi^0X$ with KOTO's to verify our simulation.

We first generate initial $K_L$ particles according to the momentum distribution of $K_L$ measured experimentally~\cite{Masuda:2015eta}, which will have a certain probability of decay in the detector. All of the $K_L$ decay modes in this paper produce two photons, and the distributions of the energies and positions of the photons are largely dependent on the probability distribution functions. For the two-body decay $K_L\to\pi^0X$, we use a uniform distribution function as in Eqs.~\eqref{eq:distpixs} and \eqref{eq:distpixv}. For the three-body decay $K_L\to\pi^0\nu\bar{\nu}$, we adopt the same distribution function as in equation~(S1) in the supplemental material of Ref.~\cite{Kitahara:2019lws}. The distribution functions of $K_L\to X\gamma\gamma$ and $K_L\to \pi^0XX$ are determined by equation~\eqref{eq:distpixxg}, and Eqs.~\eqref{eq:distpixxs} and \eqref{eq:distpixxv}, respectively. It is worth mentioning that for the decay modes $K_L\to\pi^0\nu\bar{\nu}$, $K_L\to\pi^0X$, and $K_L\to\pi^0XX$, the two photons are originated from the $\pi^0$, while for $K_L\to X\gamma\gamma$, the two photons are directly generated by $K_L$; only the former decays will lead to an invariant mass of the two photons $m_{\gamma\gamma}\simeq m_{\pi^0}$. Then the photons are captured by the CsI calorimeter in the detector, and we record their energies and positions in our simulation. In order to better simulate the detector's response to the photons, we include the energy and position resolution of the CsI calorimeter, which can be found in Ref.~\cite{Sato:2015yqa}.

\begin{figure}
\vspace{0.5cm}
\includegraphics[width=0.42\textwidth]{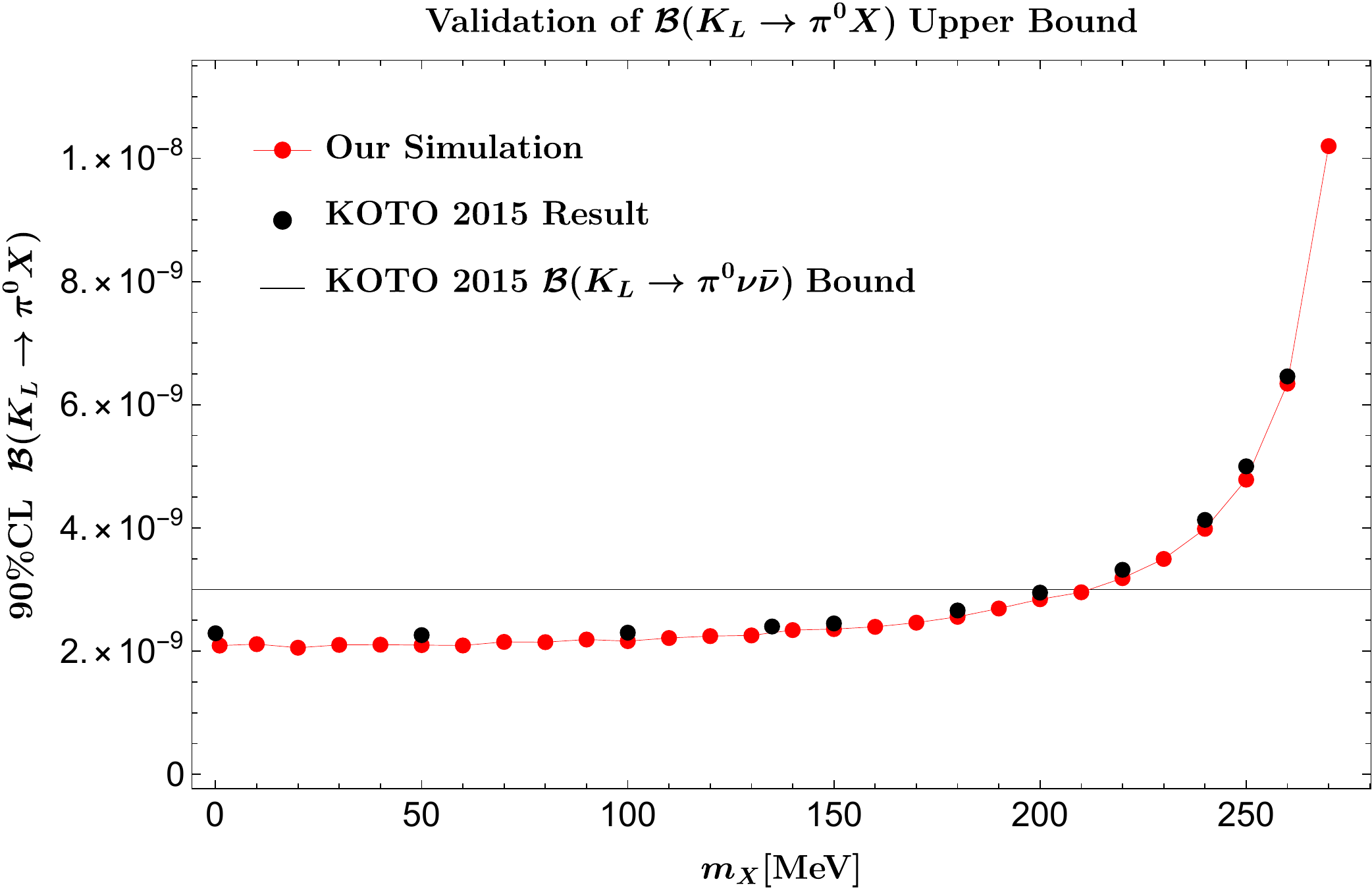} 
\caption{Validation of $\mathcal{B}(K_L\to\pi^0X)$ upper bound at $90\%$ C.L. by comparing our result (red dots) with KOTO's \cite{Ahn:2018mvc} (black dots). The horizontal line is the upper limit for $\mathcal{B}(K_L\to\pi^0\nu\bar{\nu})<3.0\times 10^{-9}$ at $90\%$ C.L..}
\label{fig:validation}
\end{figure}

By assuming that the two photons produced on the beam axis are from $\pi^0$ decay (in the interesting case of $K_L\to X\gamma\gamma$ the two photons are also required to have an invariant mass $m_{\gamma\gamma}\simeq m_{\pi^0}$ in order to fake the signals), we can reconstruct the decay location $Z_{\rm vtx}$ and $\pi^0$'s transverse momentum $P_T$ by combining the information of the photons' energies and positions. Then we use the same selection criteria as KOTO~\cite{Ahn:2018mvc} to analyze the reconstructed events. We use the signal region in KOTO's analysis of 2015 data~\cite{Ahn:2018mvc} for validation, whose result is shown in Fig.~\ref{fig:validation}, but employ the new signal region~\cite{kototalks} to analyze our decay modes. It should be noted that we do not consider shape-related cuts and veto cuts, whose efficiencies are considered the same for different processes and can be estimated as $r=\epsilon^{\rm KOTO}_{\pi^0\nu\bar{\nu}}/\epsilon_{\pi^0\nu\bar{\nu}}$, where $\epsilon_{\pi^0\nu\bar{\nu}}$ is from our rough simulation and $\epsilon^{\rm KOTO}_{\pi^0\nu\bar{\nu}}$ is the real acceptance from KOTO's systematic simulation. We calculate the real acceptance of each process by multiplying our result with $r$. After applying the reconstruction and various kinematical cuts, we finally obtain the distributions of signal events in the $Z_{\rm  vtx}-P_{T}$ plane, and the branching ratios and Wilson coefficients of some processes can be evaluated accordingly.

%


\end{document}